\documentclass[lettersize,journal]{IEEEtran}
\usepackage{amsmath,amsfonts,amsthm}
\usepackage{array}
\usepackage{textcomp}
\usepackage{stfloats}
\usepackage{url}
\usepackage{soul}
\usepackage{verbatim}
\usepackage{graphicx}
\usepackage{cite}
\usepackage[numbers, sort&compress]{natbib}
\usepackage{mathrsfs}
\usepackage{diagbox}
\usepackage{textcomp}
\usepackage{xcolor}
\usepackage{subfigure}
\usepackage{enumitem}
\usepackage{nameref}
\usepackage{multirow}
\usepackage{soul}
\usepackage{url}
\usepackage[utf8]{inputenc}
\usepackage{graphicx}
\usepackage{booktabs}
\usepackage{caption}

\usepackage[ruled,vlined]{algorithm2e}
\DeclareMathOperator*{\argmax}{arg\,max}
\DeclareMathOperator*{\argmin}{arg\,min}

\usepackage{balance}

\begin{document}
\title{Towards Adversarially Robust Recommendation from Adaptive Fraudster Detection}

\author{Yuni Lai, 
		Yulin Zhu, 
		Wenqi Fan, 
		Xiaoge Zhang, Kai Zhou
  \thanks{Yuni Lai, Yulin Zhu, Wenqi Fan, and Kai Zhou are with the Department of Computing, The Hong Kong Polytechnic University. Xiaoge Zhang is with the Department of Industrial and System Engineering, The Hong Kong Polytechnic University. Kai Zhou is the corresponding author with email: kaizhou@polyu.edu.hk.}
}

\maketitle

\begin{abstract}
    The robustness of recommender systems under node injection attacks has garnered significant attention. Recently, \textit{GraphRfi}, a GNN-based recommender system, was proposed and shown to effectively mitigate the impact of injected fake users. However, we demonstrate that \textit{GraphRfi} remains vulnerable to attacks due to the supervised nature of its fraudster detection component, where obtaining clean labels is challenging in practice. In particular, we propose a powerful poisoning attack, \textbf{MetaC}, against both GNN-based and MF-based recommender systems. Furthermore, we analyze why \textit{GraphRfi} fails under such an attack. Then, based on our insights obtained from vulnerability analysis, we design an adaptive fraudster detection module that explicitly considers label uncertainty. This module can serve as a plug-in for different recommender systems, resulting in a robust framework named \textbf{PDR}. Comprehensive experiments show that our defense approach outperforms other benchmark methods under attacks. Overall, our research presents an effective framework for integrating fraudster detection into recommendation systems to achieve adversarial robustness.
\end{abstract}


\section{Introduction}
Recommender systems (RS) are now considered an essential component of online shopping platforms like Amazon, Taobao, and eBay. By analyzing customers' historical shopping behaviors, including the items they have browsed, reviewed, or rated, RS can provide personalized product recommendations to potential customers who may be interested in them. A typical RS is built around a machine learning algorithm that operates on a bipartite graph. Specifically, the graph comprises two sets of nodes representing users and items, and the edges between them indicate the ratings that users have given to items. To generate personalized recommendations, various graph analytic techniques, such as Matrix Factorization (MF)~\cite{koren2009matrix,mehta2017review}  and Graph Neural Networks (GNNs)~\cite{wu2020graph,gao2022graph}, have been utilized to predict the missing ratings. Based on these predicted ratings, the RS recommends items to users that are likely to be of interest to them with higher predicted ratings resulting in higher recommendation priority.

Similar to other machine learning-based systems \cite{wang2023attacking,lukas2022sok}, the adversarial robustness of RS has been a topic of significant research interest. One reason is that the predictions of RS are crucial for sellers to generate profits and for users to make informed decisions, making RS a tempting target for attackers. Additionally, it is relatively easy for attackers to manipulate the graph data that RS operates on in real-world scenarios. For example, attackers can inject fake user accounts with manipulated ratings to deceive the prediction results of an RS. In fact, node injection attacks~\cite{Zhang_2021,Zhang_2020,Si_2018} have emerged as the primary form of attacks, where fake nodes, called fraudsters, are injected with carefully crafted ratings to intentionally alter the recommendation outcomes.


Against this background, various defense mechanisms \cite{wu2021fight,tang2019adversarial} have been proposed to improve the adversarial robustness of RS against node injection attacks. These defense strategies can be broadly classified into two categories: training robust models and detecting malicious fraudsters. The first approach involves training RS models with robust parameters that can accurately predict outcomes even when the input data is poisoned. Adversarial training~\cite{anelli2022adversarial} is one of the primary techniques used to achieve this goal. The second approach focuses on developing specialized techniques to detect and identify malicious fraudsters. These techniques are designed to filter out or mitigate the impact of fraudsters on the RS. It is worth noting that the two defense approaches mentioned above are not mutually exclusive. Rather, they are complementary and can be used together to further enhance the overall adversarial robustness of RS. In this paper, our focus is on the second approach with the key research question of how to effectively integrate fraudster detection into RS.

One of the most representative fraudster-detection-based methods is \textit{GraphRfi}~\cite{zhang2020gcn}.  
 At a high level, \textit{GraphRfi} innovatively attaches the GNN-based recommendation component with a fraudster detection component, which can produce the anomaly probability of each user. This probability is further used as the weight for that user in the training objective function for recommendation, such that users with a higher anomaly probability would have a lower contribution. \textit{GraphRfi} trains both components jointly in an end-to-end manner, resulting in state-of-the-art performance for robust recommendation under node injection attacks.
Overall, \textit{GraphRfi} offers a promising way of integrating fraudster detection into recommendation to achieve robustness. However, our analysis shows that \textit{GraphRfi} is still vulnerable to node injection attacks. The underlying reason, as we show in Section~\ref{section-reason}, is that the fraudster detection component relies on a supervised learning method, which in turn relies on the availability of initial user labels (i.e., fake or normal) to accurately detect anomalies. In practice, obtaining these true labels is extremely difficult, if not impossible. Even if unsupervised anomaly detection methods are used to preprocess the data and label users, the results may contain errors. This leads to noisy user labels, where some fake users are labeled as normal, causing \textit{GraphRfi} to assign large weights to these fake users due to its supervised nature. As a result, GraphRfi may malfunction and not effectively detect fraudsters, rendering it vulnerable to node injection attacks.

Thus, to address the above limitations, we propose a novel way of integrating fraudster detection into recommendation, resulting in 
a robust recommendation framework that could be applicable to widely-used recommender systems.
To this end, our first step is to conduct a thorough vulnerability analysis of fraudster-detection-based robust RS with \textit{GraphRfi} as the representative. Specifically, we design a powerful node injection attack which is formulated as a bi-level combinatorial optimization problem. We utilize a gradient-based method to solve the problem, where one of the main challenges is to compute the required gradients. We adopt the idea of meta-gradients proposed by \citet{zugner2018adversarial} that is designed to attack Graph Convolutional Networks 
(GCNs) \cite{kipf2016semi} via manipulating the graph structure (i.e., add/delete edges). Different from the attack proposed in \cite{zugner2018adversarial} that is designed for unweighted graphs, we not only need to decide the optimal injected edges between fake users and items but also the optimal rating associated with each edge. Our solution is to use a \textbf{continuous} \textit{rating probability tensor} to encode all discrete ratings. After optimization, we use discretization techniques to recover the desired ratings. We term our attack as \textbf{metaC}. Our experiments show that \textbf{metaC} is very effective in promoting targeted items even with small budgets against the robust model \textit{GraphRfi} as well as the MF-based model.

Then, based on our vulnerability analysis of \textit{GraphRfi}, 
we design a general robust recommendation framework termed Posterior-Detection Recommender (\textbf{PDR}) featured with an \textit{adaptive fraudster detection} module.
In particular, this new fraudster detection module will take label uncertainty into consideration and is jointly trained with existing recommender systems to enhance their robustness. 
Specifically, we treat the input user labels as observed but uncertain and changeable variables (\textit{priors}). We then employ an Implicit Posterior (IP) model~\cite{rolf2022resolving} to estimate the \textit{posterior} probability of the true label. Furthermore, we use a strategy to dynamically adjust the prior labels based on the estimated posterior probabilities to counter the noise. The effect is that even if the input labels are noisy, they can be properly adjusted during the training process. Consequently, the fake users (even though mislabeled as normal) would have fewer contributions to the recommendation, which makes our proposed \textbf{PDR} robust against attacks. 

We implement \textbf{PDR} with GNN-based and MF-based as the base RS, respectively, against the powerful attack \textbf{MetaC}.
Our comprehensive experiments demonstrate that \textbf{PDR} can significantly mitigate the attack effects of \textbf{MetaC} and outperforms other defense baselines.
This highlights the efficacy of our proposed adaptive fraudster detection module as a viable plug-in, which results  in a general framework \textbf{PDR} to provide adversarial robustness for recommendation.


In summary,  we investigate both the attack (\textbf{metaC}) and defense (\textbf{PDR}) approaches to comprehensively analyze the adversarial robustness of recommender systems. The main contributions of this work are summarized as follows:
\begin{itemize}
    \item We propose a new attack method \textbf{metaC} that is tested effective against both GNN-based and MF-based recommender systems. In particular, we conduct a detailed analysis of the causes of the vulnerability of a representative robust RS \textit{GraphRfi}, providing insights for properly integrating fraudster detection into RS.
    \item We propose a new robust RS framework (\textbf{PDR}). This framework features a novel dynamically adaptive fraudster detection module that significantly improves the adversarial robustness of RS under attacks.
    
    \item We conduct comprehensive experiment evaluations on two representative RS models
    to demonstrate the superior performances of both our proposed attack and defense approaches. We believe that \textbf{PDR} can serve as a framework for enhancing the robustness of RS by incorporating an anomaly detection component that can dynamically adjust the contributions of data points.
\end{itemize}

The rest of this paper is organized as follows. We summarize the related work in Section~\ref{RelatedWork}. In Section~\ref{Preliminaries}, we introduce the necessary background on our targeted recommender systems. We describe the threat model in Section~\ref{SystemSetting}. In Section~\ref{Attacks}, we present our proposed  attack method \textbf{MetaC}. Then, we propose, \textbf{PDR}, a robust recommender system framework in Section~\ref{RobustRecommendation} 
using \textit{GraphRfi} as the illustrative example. We investigate how to generalize both \textbf{MetaC} and \textbf{PDR} to the MF-based model in Section~\ref{GeneralizationMF}. We conduct extensive experiments  in Section~\ref{Experiments} to show the effectiveness of our proposed attack and defense. Finally, we conclude our findings in Section~\ref{Conclusion}.

\section{Related Works}
\label{RelatedWork}
\subsection{Attacks on recommender system} 
Injecting nodes into the recommender system is the major attacking approach as it could be easily implemented in practice.
The difficulty, however, lies in the selection of items and the ratings they give. Earlier attacks~\cite{li2016shilling,sundar2020understanding} rely on choosing filler items by heuristic rules and giving the highest/lowest ratings to the target items, depending on the goal of pushing or nuking items. However, these attacks are not effective enough 
as shown by \cite{rezaimehr2021survey} and \cite{wu2021triple,wu2021ready}.

Recently, more sophisticated methods have been proposed based on techniques such as optimization, generative models, and so on.
For instance, \cite{fang2020influence} 
proposed a method to optimize the selection of filler items using approximated gradients to attack MF-based RS. \cite{Hai2021data} train a poisoned RS model to predict the ratings for filler items. 
In addition, another line of works~\cite{christakopoulou2019adversarial,wu2021ready,wu2021triple} explore the utilization of generative models (e.g., GAN \cite{goodfellow2020generative}) to generate fake users profiles, which are injected into the system.
However, previous works mainly focus on MF-based models due to their simplicity, and could not be easily extended to GNNs-based systems. 
For example, the approximating gradients proposed in ~\cite{fang2020influence,christakopoulou2019adversarial} 
cannot be directly applied to GNNs-based models like \textit{GraphRfi}. 

\subsection{Defenses on Recommender System}
The primary method to achieve the adversarial robustness of RS is through adversarial training which has been tested effective in many other machine learning systems. ~\cite{he2018adversarial} and ~\cite{tang2019adversarial} 
perform adversarial training by adding perturbation noise to model parameters in each training iteration to improve the robustness of different target models.
\cite{wu2021fight} train a robust MF-based RS via injecting some defense users based on the calculation of influence functions.
Again, it is nontrivial to extend such an idea of defense to GNNs-based systems due to the complexity of estimating the Hessian matrix in the influence function.


A topic closely related to defense is Fraudster (or anomaly) detection, as it is natural to detect injected fake users.
Basically, anomaly detection methods \cite{burke2006classification,yu2021detecting} extract the user features according to the behavior of the users, such as the degree, rating time, and review content, and then apply classification techniques to identify the abnormal users. Due to difficulty in obtaining the label, unsupervised methods such as clustering~\cite{mehta2007unsupervised,zhang2014detection,hao2021unsupervised,zhang2021user} and semi-supervised methods \cite{wu2012hysad,cao2013shilling} are widely-used in detection. 
However, we emphasize that anomaly detection is often employed as a preprocessing step. 
Given the current under-explored status of GNNs-based RS, we aim to investigate the adversarial robustness of the representative model \textit{GraphRfi}.

\section{Preliminaries}
\label{Preliminaries}

\subsection{Recommender Systems}
A recommender system (RS)  typically operates on a weighted bipartite graph $\mathcal{G}=\left(\mathcal{U}\cup \mathcal{V},\mathcal{E}\right)$, where $\mathcal{U} = \{u_1, \cdots, u_n\}$ is a set of $n$ users, $\mathcal{V} = \{v_1, \cdots, v_m\}$ is a set of $m$ items, and the edge set $\mathcal{E} = \{e_{ij} = (u_i,v_j,r_{ij})\}$ is a collection of observed ratings with $r_{ij} \in \{1,2, \cdots, r_{max}\}$ denoting the rating from user $u_i$ to item $v_j$. 
Each user $u_i$ is also associated with a feature vector $\mathbf{x}_i$ summarizing this user's behavioral information. The task of recommendation thus amounts to predicting the weights of missing edges and recommending highly ranked items to users.

Recommender systems can be implemented using various techniques. One of the most classical methods is Matrix Factorization (MF) \cite{mehta2017review}. More recently, graph representation learning techniques, such as Graph Neural Networks (GNNs), have been increasingly utilized to improve prediction performance \cite{qiao2022tag,wu2022graph,chen2021structured,deng2022graph}. In the following, we introduce some representative models to provide background on the techniques used in recommender systems.

\subsubsection{MF-based RS} 
Matrix factorization (MF)-based recommendation models, such as SVD~\cite{sarwar2000application}, Regularized SVD~\cite{RSVD2006Simon}, and Improved Regularized SVD~\cite{paterek2007improving}, are widely used in recommendation systems due to their simplicity and effectiveness. For instance, Regularized SVD predicts missing values in the history rating matrix by decomposing it into user embedding matrix $U$ and item embedding matrix $V$. The embedding matrices $U$ and $V$ are learned by regression on existing/history ratings as follows:
\begin{equation}
    \argmin_{U,V} \sum_{\forall (u,v) \in \mathcal{E}} (r_{uv}-U_u^TV_v)^2+\beta(\lVert U \rVert + \lVert V \rVert),
\end{equation}
where the $U_u$ and $V_v$ are the embedding for user $u$ and item $v$ respectively, and $\beta$ is the regularization factor.

\subsubsection{GNN-based RS}
We use \textit{GraphRfi} as a representative to introduce GNN-based RS. 
To mitigate \textit{node injection attacks}, a robust RS \textit{GraphRfi} was introduced that combines recommendation with fraudster (i.e., fake users) detection. In particular, \textit{GraphRfi} has two essential components: a rating prediction component based on Graph Neural Networks (GNNs)  and a fraudster detection component based on Random Neural Forest (RNF).
The main idea of \textit{GraphRfi} is to treat the anomaly score of a user (from the fraudster detection component) as her weight in estimating the ratings, thus mitigating the effects of anomalous users. 
The backbone of this RS model is GNNs with attention aggregators. It encodes the discrete ratings by learnable embeddings $e_r\in \mathbb{R}^d$, $r \in \{1,2,\cdots,r_{max}\}$, where $d$ is the dimension. For each user/item, it concatenates the rating embedding with user/item embedding, and then two single-layer GNNs are employed to learn the user's and item's representation,
\begin{equation}
\resizebox{.99\linewidth}{!}{$
            \displaystyle
\begin{aligned}
&z_u=ReLU(W_1 \cdot Agg(\{\text{MLP}(x_v \oplus e_{r_{uv}}), \forall v \in \mathcal{N}(u)\})+b_1),\nonumber \\
&z_v=ReLU(W_2 \cdot Agg(\{\text{MLP}(x_u \oplus e_{r_{uv}}), \forall u \in \mathcal{N}(v)\})+b_2),\nonumber 
\end{aligned}
$}
\end{equation}
where 
$x_u$/$x_v$ is initial user/item embedding,
$\oplus$ is concatenation, and $Agg(\cdot)$ is attention aggregation function, 
$$Agg({h_k,\forall k\in \mathcal{N}(s)})=\sum_{k\in\mathcal{N}(s)} \alpha_{ks}h_k ,
$$
and the $\alpha_{ks}$ is the weight learned by attention layer,
$$
a_{ks}= W_4 \cdot \sigma (W_3\cdot (h_k\oplus z_s)+b_3)+b_4,
$$
$$
\alpha_{ks}=\frac{exp(a_{ks})}{\sum_{k'\in\mathcal{N}(s)} exp(a_{k's})}.
$$

In other words, GNNs are used to learn the embeddings of both users and items denoted as $z_u$ and $z_v$, which are further used to compute the predicted rating $r_{uv}'$ from user $u$ to item $v$ through a multi-layer perceptron (MLP): 
$$
r_{uv}'=W_5 \cdot \text{MLP}(z_u \oplus z_v),
$$
where $W_i$, $b_i$ are the learnable parameters, $\mathcal{N}(s)$ is the neighbor set of node $s$.

Given the user embedding $z_u$ learned by GNNs, a classifier (i.e, Random Neural Forest \cite{biau2019neural}) is used to estimate the probability that a user $u$ is normal, denoted as $\mathbb{P} \left[ y=0 | z_{u}, \theta \right]$, where $\theta$ is the model parameter, and $y=0$ indicates that a user is normal. 
Finally, the prediction and detection components are jointly trained in an \textit{end-to-end} manner by minimizing the following loss function consisting of two parts:
\begin{equation}
\setlength{\jot}{2mm}
\begin{aligned}
\label{eqn-loss-function}
\mathcal{L}(\theta,\mathcal{G}) &= \mathcal{L}_{\text{rating}} + \lambda \cdot \mathcal{L}_{\text{fraudster}} \\
&={\textstyle\frac{1}{\left| \mathcal{E} \right| } \sum_{\forall (u,v) \in \mathcal{E} }} \mathbb{P} \left[ y=0 | z_{u} , \theta \right] \cdot \left( r_{uv}^{\prime}-r_{uv}\right)^2  \\
&+ {\textstyle \frac{1}{\left| \mathcal{U} \right| } \sum_{\forall u \in \mathcal{U},y_u \in \mathcal{Y} }} \left( -\log \mathbb{P} \left[ y=y_u | z_{u} , \theta \right]\right), 
\end{aligned}
\end{equation}
where $\mathcal{L}_{\text{rating}}$ summarizes the weighted mean squared error of rating and $\mathcal{L}_{\text{fraudster}}$ is the cross-entropy loss for anomaly detection with $y_u$ denoting the ground-truth label of user $u$.


The probability $\mathbb{P} \left[ y=0 | z_{u} , \theta \right]$ serves as the weight for user $u$. As a result, a user with a high anomaly score (i.e., $1- \mathbb{P}$) contributes less to the prediction, which can enhance the robustness of recommendation under node injection attacks. We can also notice that the fraudster detection component is supervised in nature as the ground-truth labels are required during training; later, we will show the defects of this design by designing a powerful attack \textbf{MetaC}.


\subsection{Posterior Estimation}
In Bayesian statistics, Maximum A Posterior (MAP) estimate is a method for estimating an unknown quantity based on observed data and prior knowledge. It is obtained by finding the distribution that maximizes the likelihood function incorporated with a prior distribution.
Suppose that there are $n$ samples with feature $z_i, i\in \{1,\cdots,n\}$, and each sample has a corresponding unknown variable $l_i$. If the prior probability for each $l_i$ is defined as $p(l_i), i\in \{1,\cdots,n\}$, and the observation is $z_i$, the posterior probability $q(l_i|z_i)$ based on the prior $p(l_i)$ and observation $z_i$ can be estimated by maximizing the log-likelihood. According to the negative evidence lower bound (ELBO), the inequality of the negative log-likelihood regarding all observed data $z_i$ is as follows:
\begin{align}
    &-\sum_{i} log(p(z_i))\nonumber\\
    &\leq \underbrace{- \sum_{i} q(l_i=c|z_i) log(\frac{p(z_i|l_i)p(l_i=c)}{q(l_i=c|z_i)})}_{F},
\end{align}
where the term $F$ is also known as free energy, and minimizing $F$ leads to maximization of the log-likelihood $\sum_{i} log(p(z_i))$. 

It is common to have coarse and imprecise labels in computer vision tasks, such as segmentation, since high-resolution labels are usually hard to obtain. Implicit Posterior model (IP) \cite{rolf2022resolving} was first employed to resolve this label uncertainty problem, and it treats uncertain labels as priors and features $z_i$ as observed data. To estimate the posterior probability of the uncertain label, $q(l_i|z_i)$ can be parameterized by a neural network. Then the free energy $F$ is equivalent to the loss function $\mathcal{L}_{IP}$ that guide the optimization of posterior $q_{\theta}(l_i=c|z_i)$:
\begin{equation}
\mathcal{L}_{IP}\!=\!\sum_{i} \! \left( q_{\theta}(l_i=c|z_i) \\
\log \frac{\sum\limits_{j} q_{\theta}(l_j=c|z_j)}{p(l_i=c)}\right).\nonumber
\end{equation}
where $\theta$ denotes the trainable parameters of neural network, and $q_{\theta}(l_i=c|z_i) := q(l_i=c|z_i;\theta)$. The minimization of the IP loss ($\mathcal{L}_{IP}$) leads to the maximization of log-likelihood regarding all observed data $z_i$.

\section{Recommendation in Adversarial Environemnt}
\label{SystemSetting}
In this section, we introduce the adversarial environment that a recommender system operates in. We consider an adversarial environment consisting of an attacker and a defender, where the attacker launches node injection attacks against a target RS while the defender (e.g., system administrator) aims to effectively run the RS in the presence of this proactive attacker. Below, we specify the goal, knowledge, and ability of both the attacker and defender, respectively.
\subsection{Attacker}
We consider an attacker whose goal is to promote a set of target items $\mathcal{T} \subset \mathcal{V}$. More specifically, the attacker aims to increase the probability that a target item $v_t \in \mathcal{T}$ appears in the top-$k$ recommendation lists of target users. Based on Kerckhoffs’s principle \cite{Kerckhoffs1883cryptographie}, we
assume a worst-case scenario where the attacker has full knowledge of the target RS, including the data (i.e., the clean graph $\mathcal{G}$) and the recommendation algorithm. To achieve the malicious goal, the attacker is able to inject a set of fake users $\mathcal{U}'$ as well as some ratings (i.e., edges $\mathcal{E}'$ between $\mathcal{U}'$ and $\mathcal{V}$), resulting in a manipulated graph $\mathcal{G}' = (\mathcal{U}\cup \mathcal{U}'\cup \mathcal{V},\mathcal{E}\cup \mathcal{E}')$. To constrain the attacker's ability, we assume that there are at most $H$ fake users (i.e., $|\mathcal{U}'| \leq H$), and each fake user can give at most $B$ ratings. After the attack, the defender observes the manipulated graph $\mathcal{G}'$, from which the RS is trained and tested; this attack falls into the category of data poisoning attacks.

\subsection{Defender}
The defender can only observe the poisoned graph $\mathcal{G}'$ instead of the clean one $\mathcal{G}$. The goal of the defender is to train a robust RS over $\mathcal{G}'$ that can mitigate the malicious effects of the injected fake users. Specifically, it is expected that with the robust RS, the target items would not be significantly promoted. We note that the defender does not know which the target items are, and we only use such information for evaluation purposes. In practice, it is common for the defender to run anomaly detection systems to filter out fraudsters \textit{before training}. To reflect this fact, we assume that the defender can identify a fraction $\tau$ ($0\% \leq \tau \leq 100 \%$) of fake users reliably. This parameter $\tau$ indicates the defender's prior knowledge about the attacks; however, we emphasize that our proposed robust RS works even when $\tau = 30\%$. 

\section{Attacks against Existing RS}
In this section, we use the GNN-based robust recommender system \textit{GraphRfi} as the example to illustrate our attack. We show that our attack can be extended to MF-based RS in Section~\ref{GeneralizationMF}.
\label{Attacks}
\subsection{Attack Formulation}
We begin by quantifying the attacker's malicious goal. Recall that the attacker aims to promote a set of target items $\mathcal{T}$, a commonly used metric to measure the effectiveness of attack for an item is the \textit{hit ratio}. Specifically, a hit ratio for an item $v$ with parameter $k$ (denoted as $HR@k (v, \mathcal{G}, \theta)$) is the percentage of users whose top-$k$ recommendation list includes that item. Note that we make explicit the dependency of the hit ratio of $v$ on the graph $\mathcal{G}$ and the trained model parameter $\theta$.
Thus, we can use an adversarial objective function $F_{adv}(\mathcal{G}, \mathcal{T}, \theta) = \frac{1}{\mathcal{T}}\sum_{v \in \mathcal{T}} HR@k (v, \mathcal{G}, \theta)$, the average hit ratios of those target items, to quantify the attacker's goal.

Poisoning attacks against recommendation then amounts to finding the optimal poisoned graph $\mathcal{G}'$ to maximize the adversarial objective function. It can be formulated as a \textit{bi-level} optimization problem, where in the outer level the attacker optimizes the objective over the graph $\mathcal{G}'$ while in the inner level, the model parameter $\theta$ is optimized though minimizing the training loss, also depending on $\mathcal{G}'$. Mathematically, a poisoning attack is formulated as:
\begin{equation}
\setlength{\jot}{0.01mm}
\begin{aligned}
\label{eqn-org-problem}
\max_{\mathcal{G}'}\quad &F_{adv}(\mathcal{G}, \mathcal{T}, \theta^*) = \frac{1}{\mathcal{T}}\sum_{v \in \mathcal{T}} HR@k (v, \mathcal{G}', \theta^*)\\
\text{s.t.} \quad &\theta^* = \argmin_{\theta} \mathcal{L}(\theta, \mathcal{G}'), \quad
\mathcal{G}' = \mathcal{G} \cup \mathcal{U}' \cup \mathcal{E}',\\
&|\mathcal{U}'| \leq H,\quad d(u') \leq B, \quad \forall u' \in \mathcal{U}',
\end{aligned}
\end{equation}
where we use $\mathcal{G}' = \mathcal{G} \cup \mathcal{U}' \cup \mathcal{E}'$ to denote that $\mathcal{G}'$ is obtained by injecting a set of fake users $\mathcal{U}'$ and edges $\mathcal{E}'$ into $\mathcal{G}$, $|\mathcal{U}'| \leq H$ requires that at most $H$ fake users are injected, and the degree constraint of fake user $d(u') \leq B$ requires that each fake user can give at most $B$ ratings.

\subsection{Attack Method}
\subsubsection{Reformulation of attack}
The major challenges in solving the above optimization problem are the discrete search space and the exponential growth of candidate edges in $\mathcal{G'}$: the attacker needs to determine which items to rate as well as the specific discrete ratings (e.g., scale $1$ to $5$). We thus use a series of techniques to approximate this discrete optimization problem. First, we use a continuous probability vector $\hat{\mathbf{r}}= (p_1, p_2, \cdots, p_{r_{max}})$ (e.g., $r_{max}=5$) to encode a discrete rating, denoting a user will give a rating $ l \in \{1,2, \cdots, r_{max}\}$ with probability $p_l$. Then, we assume that the injected users will initially connect to \textit{all} items. Thus, the attacker's behavior is now fully captured by a continuous rating tensor $\hat{\mathbf{R}} \in [0,1]^{| \mathcal{U}^\prime | \times | \mathcal{V} | \times r_{max}}$. We denote the manipulated graph as $\hat{\mathcal{G}}= \mathcal{G} \cup \mathcal{U}' \cup \hat{\mathbf{R}}$. Another difficulty comes from the non-differentiability of the objective function, in particular, the hit ratios. Thus, we use a sum of $softmax$ function ratios \cite{tang2020revisiting} to approximate $F_{adv}(\mathcal{G}, \mathcal{T}, \theta^*)$ as below:
\begin{equation}
\mathcal{L}_{adv}( \hat{\mathcal{G}} , \mathcal{T},\theta^*)= - \sum_{t \in \mathcal{T}} \sum_{u \in \mathcal{U}} \log ( \frac{\exp{( r_{ut}') }}{\sum_{v \in \mathcal{V}} \exp{( r_{uv}') }}),
\end{equation}
where $r_{uv}'$ denotes the predicted rating from user $u$ to $v$. Basically, this function measures the fraction of the ratings for targeted items over the ratings for all items. Now, the optimization problem defined in Eq. (\eqref{eqn-org-problem}) is recast as:
\begin{equation}
\setlength{\jot}{0.01pt}
\begin{aligned}
\label{eqn-reformulate}
\min_{\hat{\mathbf{R}}} \quad &\mathcal{L}_{adv}( \hat{\mathcal{G}} , \mathcal{T},\theta^*)\\
s.t. \quad & \theta^*= \arg\min_{\theta} \mathcal{L}(\theta, \hat{\mathcal{G}})
, \, \hat{\mathcal{G}}= \mathcal{G} \cup \mathcal{U}' \cup \hat{\mathbf{R}},\\
&  \sum_{l} \hat{\mathbf{R}}_{i,j,l} =1(\forall i,j), \, \hat{\mathbf{R}} \in\left[0,1 \right]^{ |\mathcal{U}'| \times |\mathcal{V}| \times r_{max}}. 
\end{aligned}
\end{equation}

To utilize the continuous rating probability tensor $\hat{R}$, 
we design the loss function during attack optimization as below: 
\begin{equation}
\label{loss-for-tensor}
    \mathcal{L}(\theta, \hat{\mathcal{G}})=\sum_{\forall (u,v) \in \mathcal{E} } \max_{l}\hat{\mathbf{R}}_{uvl} \cdot ( r_{uv}^{\prime}-\argmax_l \hat{\mathbf{R}}_{uvl})^2,
\end{equation}
where the term $\argmax_l \hat{\mathbf{R}}_{uvl}$ is used to find the maximum index of the probability vector as the ground truth, and $\max_{l}\hat{\mathbf{R}}_{uvl}$ is the associated maximum probability value. Rating vectors with higher maximum probability will have a higher contribution to the RS training loss during the attack optimization.

\subsubsection{Optimization method}
Now, we describe the method to solve problem \eqref{eqn-reformulate} to obtain the (sub-)optimal continuous rating tensor $\hat{\mathbf{R}}$, from which we derive the discrete ratings that satisfy all the constraints. We alternately update the inner objective function $\mathcal{L}(\cdot)$ with respect to $\theta$ for $K$ steps and update the outer objective function $\mathcal{L}_{adv}(\cdot)$ with respect to $\hat{\mathbf{R}}$ for one step, where $K$ is a hyper-parameter. 
However, the central challenge lies in computing the gradients of $\mathcal{L}_{adv}(\cdot)$ with respect to $\hat{\mathbf{R}}$ because $\theta^*$ itself is obtained through an optimization process depending on $\hat{\mathbf{R}}$. 
 We adapt the idea of \textit{approximating meta gradients} \cite{zugner2018adversarial}
 to compute the required gradients. 
 In detail, we sum the gradients $\nabla_{\hat{\mathbf{R}}} \mathcal{L}_{adv}( \hat{\mathcal{G}}, \mathcal{T},\theta^{t})$ during the $K$ steps of inner model updates as the approximation, termed as meta-gradients:
\begin{equation}
\label{meta-gradient}
	\nabla_{\hat{\mathbf{R}}}^{meta} \mathcal{L}_{adv} \approx \sum_{t=1}^{K} \nabla_{\hat{\mathbf{R}}} \mathcal{L}_{adv}( \hat{\mathcal{G}} , \mathcal{T},\theta^{t}).
\end{equation}

\begin{figure*}[!ht]
\centering
\includegraphics[width=18cm]{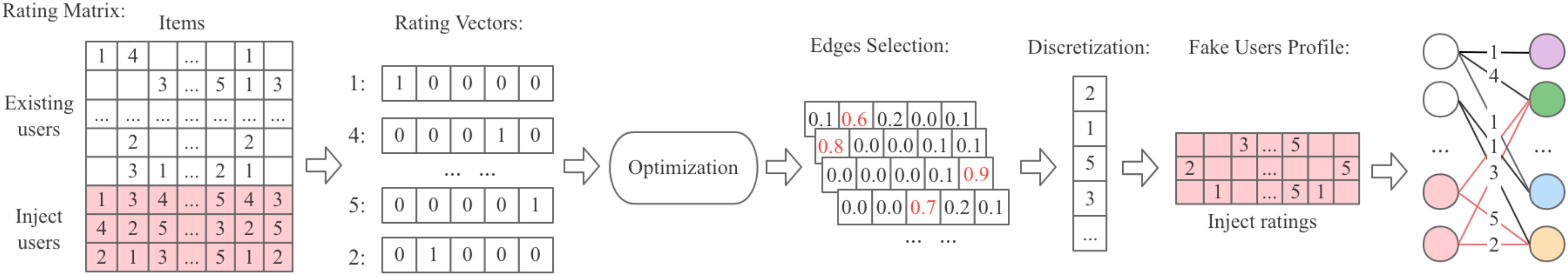}
\caption{Illustration of poisoning attack \textbf{MetaC}.}
\label{fig:metaC_framework}
\end{figure*}

As a result, we can update $\mathcal{L}_{adv}$ for one single step in the outer layer based on meta-gradients:
\begin{equation}
\hat{\mathbf{R}}^{t+1} = \hat{\mathbf{R}}^{t} - \eta_2 \cdot \nabla_{\hat{\mathbf{R}}}^{meta} \mathcal{L}_{adv},
\end{equation}
where $\eta_2$ is the learning rate. 
After each update of $\hat{\mathbf{R}}$, we conduct zero-one scaling and normalization to ensure that the entries of $\hat{\mathbf{R}}$ always stay within the range of $[0,1]$ and sum to $1$ while gradient updates might break these constraints. Specifically, we employ the commonly used $0$-$1$ normalization method on all entries (for $\forall i,j$):
\begin{equation*}
\hat{\mathbf{R}}_{i,j,:} =\frac{{\hat{\mathbf{R}}}_{i,j,:} - \min({\hat{\mathbf{R}}}_{i,j,:})}{\max({\hat{\mathbf{R}}}_{i,j,:}) - \min({\hat{\mathbf{R}}}_{i,j,:})},
\, \hat{\mathbf{R}}_{i,j,:} =\frac{\hat{\mathbf{R}}_{i,j,:}}{\sum_{l=1}^{L} \hat{\mathbf{R}}_{i,j,:}}.
\end{equation*}

These operations ensure that the entries of $\hat{\mathbf{R}}$, which is $\hat{\mathbf{R}}_{i,j,:}\in \mathbb{R}^{r_{max}}$, can be interpreted as the probabilities of adding specific ratings during optimization. We can then iteratively update $\hat{\mathbf{R}}$ and $\theta$ until the loss $\mathcal{L}_{adv}$ diminishes to an acceptable level.




After the optimization, we will need to discretize the ratings and simultaneously pick $B$ (budget) ratings for each fake user. 
Since the highest value of probabilities $\max_{l}\hat{\mathbf{R}}_{uvl}$ are served as weights in the training loss in Eqn~\eqref{loss-for-tensor}, we discretize each rating vector $\hat{\mathbf{R}}_{ij}$ as $(l,p_l)$ and pick top-$B$ ratings by ranking the $p_l$ in descending order as the injected ratings, where $p_l = \max_{l} \hat{\mathbf{R}}_{ij}$ and the discrete rating $l$ is obtained by the corresponding index of $p_l$. In doing so, we have tackled the problem of simultaneously determining the optimal edges and ratings.

We further summarize the process of obtaining continuous rating tensor $\hat{\mathbf{R}}$ through alternate iteration in Algorithm \ref{alg-attack}, and Fig.~\ref{fig:metaC_framework} shows the framework of \textbf{MetaC}. Firstly, we initialize the $\hat{\mathbf{R}}$ by sampling ratings from the normal distribution $N(\mu,\sigma^2)$, where $\mu$ and $\sigma^2$ are the mean and variance of existing history ratings. Secondly, we do inner training (the RS model) for $K$ steps with fixed $\hat{\mathbf{R}}$. Thirdly, we update $\hat{\mathbf{R}}$ by approximated meta-gradient $\nabla_{\hat{\mathbf{R}}}^{meta} \mathcal{L}_{adv}$ aiming to optimize adversarial attack loss $\mathcal{L}_{adv}$, and scale the vectors in $\hat{\mathbf{R}}$ to satisfy the probability constraints. We update the $\theta$ and $\hat{\mathbf{R}}$ alternately for $T_{train}$ epochs.
\begin{algorithm}[htbp]
	\caption{\textbf{MetaC} Poisoning Attack}
	\label{alg-attack}
	\KwIn{ Initiated rating tensor $\hat{R}$; Total training epochs $T_{train}$; Inner training steps $K$}
	\KwOut{ The optimized rating tensor $\hat{R}$}
	\BlankLine
	\For{$t\leftarrow 0$ \KwTo $T_{train}$}{
		\For{$k\leftarrow 0$ \KwTo $K$}{
			\emph{$\theta^{(tK+k+1)} = \theta^{(tK+k)}-\eta_1 \nabla_{\theta^{(tK+k)}} \mathcal{L}(\theta^{(tK+k)}, \hat{\mathcal{G}})$}
		}
		\emph{$\hat{R}^{t+1} = \hat{R}^{t} - \eta_{2} \cdot \nabla_{\hat{\mathbf{R}}}^{meta} \mathcal{L}_{adv}$}\;
		\emph{$\hat{R}_{i,j,:} =\frac{{\hat{R}}_{i,j,:} - min({\hat{R}}_{i,j,:})}{max({\hat{R}}_{i,j,:}) - min({\hat{R}}_{i,j,:})}$}\;
		\emph{$\hat{R}_{i,j,:} =\frac{\hat{R}_{i,j,:}}{\sum_{l=1}^{L} \hat{R}_{i,j,:}}$}
	}
	\Return{$\hat{R}$}
\end{algorithm}

In summary, we formulate the node injection poisoning attack as a bi-level optimization problem that is solved by an alternate iteration method in the continuous domain. Through discretization, we finally obtain the injected users associated with ratings, i.e., a poisoned graph $\mathcal{G}'$. We term this attack method as \textbf{MetaC}. We note that, by Kerckhoffs's principle in security, we consider a worst-case scenario when designing \textbf{MetaC}, where the attacker has full knowledge of the RS and some additional attack design goals, such as unnoticeability, are not considered. We then aim to design robust RS against this strong attack in the worst case.

\subsection{Vulnerability analysis}
\label{section-reason}

The underlying reason that \textit{GraphRfi} fails against our proposed poisoning attack is that its anomaly detection component adopts a \textit{supervised} learning approach. As a result, if a user is labeled as normal (even if it is actually fake), supervised learning will eventually assign a small anomaly score to it as the training process continues. That is, fake users that are labeled as normal would still have strong malicious effects on the prediction.

\textit{We conduct comprehensive experiments to demonstrate this phenomenon}. Specifically, we classify the users into four types. Type I and Type II users are normal and anomalous users inherently existing in the graph, respectively, and the defender knows their labels reliably. Among the injected fake users, a fraction of $\tau$ users (denoted as Type III) are determined as anomalous with high confidence by the defender and thus are labeled as abnormal. The rest of the fake users (denoted as Type IV) are labeled as normal. We emphasize that the parameter $\tau$ is a variable that reflects the defender's ability to identify abnormal users from the collected data, and in practice, it is not uncommon that $\tau$ is low especially when there are several effective stealthy node injection attacks against recommender systems.
We observed the anomaly scores for those four types of users during the whole training process of \textit{GraphRfi}, shown in Fig.~\ref{fig-score-no-defense-yelp}. We can observe that Type II and Type III users (labeled as fake) have very high anomaly scores during training. In comparison, the anomaly scores of Type IV users (fake but labeled as normal) keep decreasing as training continues and eventually approach to those of Type I users (normal).

In summary, the insufficient ability of a defender to filter out fake users (which is common) resulted in highly noisy user labels and further caused the supervised anomaly detection component to assign low anomaly scores (i.e., large weights of contribution) to \textit{evaded} fake users, which finally left \textit{GraphRfi} vulnerable to poisoning attacks. This crucial observation also guides us in designing robust RS.

\section{Robust Recommendation under Attack}
\label{RobustRecommendation}
\begin{figure*}[htbp]
\centering
\includegraphics[width=13cm]{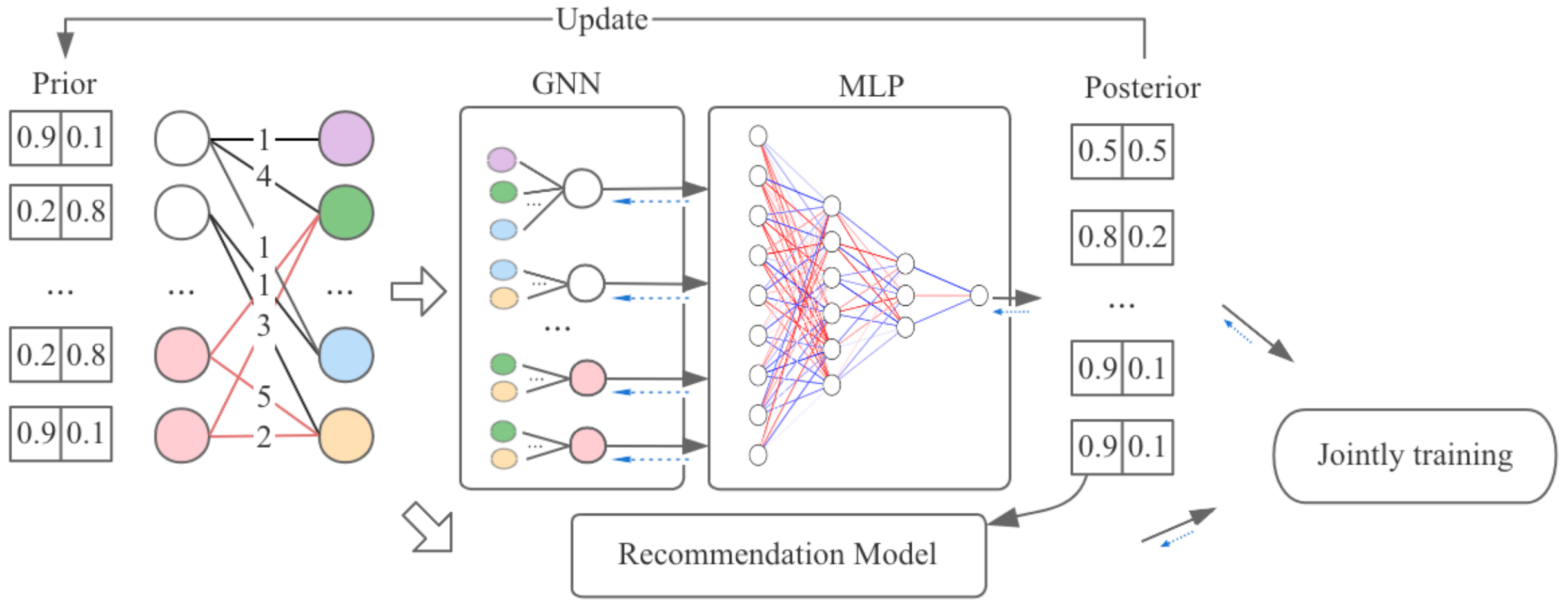}
\caption{Robust recommendation framework \textbf{PDR}.}
\label{fig:PDR_framework}
\end{figure*}

In this section, we introduce our framework \textbf{PDR} using \textit{GraphRfi} as the illustrative example. We show that \textbf{PDR} can also be extended to MF-based RS in Section~\ref{GeneralizationMF}.
\subsection{Framework}
Our previous analysis shows that the inability to label all fake users correctly causes the failure of \textit{GraphRfi}. In this section, we propose techniques to resolve this issue with the goal of building a robust recommender system termed Posterior-Detection Recommender system (\textbf{PDR}). 


Figure~\ref{fig:PDR_framework} presents the framework of \textbf{PDR}, which consists of two components: a recommendation model and an adaptive fraudster detection model. The fraudster detection model starts by assigning a prior anomaly probability to users based on the given noisy label. Then, an anomaly detection module (i.e., a GNN layer and an MLP) is used to estimate the posterior probability based on the history rating graph. If the RS is a GNN-based model (e.g., \textit{GraphRfi}), we can use the user embedding provided by the GNN in the RS directly. The detection model and recommendation model are jointly trained, and the posterior probability provided by the detection model is served as weights of users in the RS. During the training process, we adjust the prior based on the posterior. Next, we introduce the robust framework in detail.

At a high level, our anomaly detection component combines two procedures: \textit{posterior probability estimation} and \textit{dynamic label adjustment}. Specifically, we assume that a label associative with a user is a \textit{variable} instead of being fixed. We turn the given noisy labels into soft labels as priors and the \textit{unknown} true labels as latent variables. Thus, given the priors and observations (i.e., user embeddings, history rating graph), we can use a model to estimate the posterior probabilities of the true labels. Then, based on the estimated posterior probabilities, we use a strategy to dynamically adjust the soft labels (priors) during the end-to-end training process in order to estimate the posterior probabilities more accurately. 
Below, we articulate the details of the two procedures.

\subsection{Posterior probability estimation} 
We aim to estimate the true labels based on the noisy labels and observation $z_u$ (the user embedding learned by GNNs in RS). We define the true label of user $u$ as a latent variable $l_u \in C = \{f,n\}$, where $f$ and $n$ represent fake and normal, respectively.
The prior probability of this label $l_u$ is represented as a two-dimensional vector $p(l_u) = [p(l_u = f), p(l_u = n)]$, which is user-specific prior.
To take into account that the given labels are noisy, we initialize the prior probabilities as follows. For a user with a given label $f$, we set $p(l_u) =[1- p_0,p_0]$ (instead of $[1,0]$), where $p_0$ is the probability that a user labeled with fake is actually normal. Similarly, for a normal user, we set $p(l_u) = [p_1, 1- p_1]$, where $p_1$ is the probability that a user labeled with normal is actually fake. We note that $p_0$ and $p_1$ are hyper-parameters of the system that depend on the anomaly detection system used to preprocess the data.

We further denote the posterior probability of the true label as $q(l_u|z_u)=[q(l_u=f),q(l_u=n)]$. We adopt the Implicit Posterior (IP) model \cite{rolf2022resolving} to estimate the posterior probability $q(l_u|z_u)$ based on the prior $p(l_u)$ and $z_u$.
We parameterize $q(l_u=c|z_u)$ by a neural network: $q_{\theta}(l_u=c|z_u)$, where $\theta$ represents the trainable parameters. To obtain a more reliable posterior, we employ the $sofmax$ function with temperature scaling. This function produces a smoother probability when temperature $T>1$, which prevents the model from becoming overconfident \cite{keren2018calibrated}. The formula is as follows,
\begin{equation} softmax_T(x_i)=\frac{exp(\frac{x_i}{T})}{\sum_j exp(\frac{x_j}{T})}.
\end{equation}
We use this loss $\mathcal{L}_{IP}$ to train our fraudster detection component which will output the estimated posterior probability $q_{\theta}(l_u| z_u)$ for each user $u$, and it serves as the weight in $\mathcal{L}_{\text{rating}}$. To integrate the IP model in the training of RS, we substitute the $\mathcal{L}_{\text{fraudster}}$ in Eqn~\eqref{eqn-loss-function} with the IP loss:
\begin{equation}
\setlength{\jot}{0.01pt}
\begin{aligned}
\label{PDR-loss-function}
&\mathcal{L}(\theta,\mathcal{G}') = \mathcal{L}_{\text{rating}} + \lambda 
 \cdot \mathcal{L}_{IP},\\
 &\mathcal{L}_{\text{rating}}={\textstyle \frac{1}{\left| \mathcal{E} \right| }} \sum_{\forall (u,v) \in \mathcal{E} } q_{\theta}(l_u=n|z_u) \cdot \left( r_{uv}^{\prime}-r_{uv}\right)^2.
\end{aligned}
\end{equation}

\begin{algorithm}[h]
	\caption{\textbf{PDR} Defense Framework}
	\label{alg-defense}
        \SetKwProg{Fn}{Def}{:}{}
	\KwIn{Poisoned Graph $\mathcal{G}'$; Total training epochs $T_{train}$; Threshold $a_0$; Update rate $\alpha$; Update interval $c_1$, $c_2$; Observed label $L$, Prior parameters $p_0$,$p_1$.}
	\KwOut{Learned RS model parameter $\theta^*$.}
	\BlankLine
        \Fn{{$clip(x,a,b)$}}{
            \emph{\quad if $x<a$:}\\
            \emph{\quad \quad $x=a$}\\
            \emph{\quad if $x>b$:}\\
            \emph{\quad \quad $x=b$}\\
            \KwRet $x$\;
        }
        \Fn{{$set\_prior(L,p_0,p_1)$}}{
        \For{$l$ in $L$}{
            \emph{\quad if $l=fake$:}\\
            \emph{\quad \quad $p(l_u |z_u) = [1- p_0,p_0]$}\\
            \emph{\quad if $l=normal$:}\\
            \emph{\quad \quad $p(l_u |z_u) = [p_1, 1- p_1]$}\\
            }
            \KwRet $p(l_u |z_u)$\;
        }
        \text{$set\_prior(L,p_0,p_1)$}\\
	\For{$t\leftarrow 1$ \KwTo $T_{train}$}{
            \emph{$\theta^{t+1} = \theta^{t}-\eta_1 \nabla_{\theta^t} \mathcal{L}(\theta^t,\mathcal{G}')$}\\
            \If{AUC $>$ $a_0$}{
                \begin{equation*}
		p(f_u)^{t+1}\! = \!
		\begin{cases}
                \text{if \,} q(f_u)^{t} < c_1:\\
			\quad (1-\alpha) p(f_u)^{t} - \alpha (1-q(f_u)^{t})\\
                \text{if \,} q(f_u)^{t} >c_2:\\
                \quad (1-\alpha) p(f_u)^{t} + \alpha q(f_u)^{t}\\
                \text{otherwise}:\\
			\quad p(f_u)^{t}
		\end{cases}
            \end{equation*}
            \emph{$q(f_u)^{t}=clip(q(f_u)^{t},0,1)$}
            \emph{$q(n_u)^{t}=1-q(f_u)^{t}$}
            }
	}
	\Return{$\theta^*=\theta^{t+1}$}
\end{algorithm}

\subsection{Dynamic label adjustment} We use another technique to estimate $q(l_u| z_u)$ more accurately. 
We observed in our experiments that as the training continues, the posterior probabilities learned by neural networks will eventually approach to the priors, probably due to the over-fitting of neural networks. To address this, we will use the highly confident posteriors to correct the errors (noise) in the priors. In other words, we will update a soft label (prior) if the corresponding posterior is of high confidence.

Specifically, we will update the soft labels in iterations along with the training process. For ease of presentation, we use $p(f_u)^t$, $q(f_u)^t$, $p(n_u)^t$, and $q(n_u)^t$ as the simplicity of $p(l_u = f)$, $q(l_u = f)$, $p(l_u = n)$, and $q(l_u = n)$ in the $t$-th iteration, respectively. We update $p(f_u)$ according to the following strategy:
\begin{equation}
		p(f_u)^{t+1}\! = \!
		\begin{cases}
			(1-\alpha) p(f_u)^{t} - \alpha (1-q(f_u)^{t}),&\!q(f_u)^{t} < c_1\\
			(1-\alpha) p(f_u)^{t} + \alpha q(f_u)^{t},&\!q(f_u)^{t} >c_2\\
			p(f_u)^{t},&\!\text{otherwise.}\nonumber
		\end{cases}
\end{equation}
Basically, we use intervals $[0,c_1]$ and $[c_2,1]$ to determine whether the estimation of $q(f_u)^t$ is confident or not. In particular, if $q(f_u)^t$ is higher than an upper-threshold $c_2$, we increase its prior probability $p(f_u)^{t+1}$ to $(1-\alpha) p(f_u)^{t} + \alpha q(f_u)^t$, where $0<\alpha<1$ is an update rate that controls the adjustment speed (i.e., the effect of $p(f_u)^{t}$ is discounted by $\alpha$). Similarly, if $q(f_u)^t$ is smaller than a lower-threshold $c_1$, we decrease $p(f_u)^{t+1}$ to $(1-\alpha) p(f_u)^{t} - \alpha (1-q(f_u)^t)$. We clip the $p(f_u)^{t+1}$ to [0,1] if it exceeds 0 or 1, and we set $p(n_u)^{t+1}=1-p(f_u)^{t+1}$.
We apply this dynamic label adjustment after the detection AUC (Area Under Curve) on the training set first reaches $a_0$ that the model has a good performance but before over-fitting, where $0.5<a_0<1$ is a hyper-parameter.
We further summarize the whole training process of \textbf{PDR} in Algorithm~\ref{alg-defense}.

\section{Generalization to MF-based RS}
\label{GeneralizationMF} 
In this section, we demonstrate that our attack and defense approaches can be applied to MF-based RS with minor modifications.

\subsection{\textbf{MetaC}} 
Different from the GNNs-based model, where $\mathcal{L}_{adv}(\hat{\mathcal{G}},\mathcal{T},\theta^{t})$ is dependent on the input $\hat{\mathbf{R}}$ when given $\theta^t$, the prediction of MF-based model only rely on $\theta^t=\{U,V\}$. This actually makes the computation of the gradients easier for MF-based RS due to its simplicity. Specifically, we can apply the meta-gradient (\emph{instead of approximating}) to directly compute $\nabla_{\hat{\mathbf{R}}}^{meta} \mathcal{L}_{adv}$.
Briefly, $\mathcal{L}_{adv}$ is depended on $\theta^K$ (obtain by $K$ steps of inner model iteration), and each $\theta^{t+1}$ depends on $\hat{\mathbf{R}}$ and $\theta^t$ during the training, so the gradient $\nabla_{\hat{\mathbf{R}}}^{meta} \mathcal{L}_{adv}$ can be traced back to the each iteration of $\theta^t$ as follow,
\begin{equation}
\setlength{\jot}{0.01pt}
\begin{aligned}
\nabla_{\hat{\mathbf{R}}}^{meta} \mathcal{L}_{adv}=&\nabla_{r'_{\theta^K}} \mathcal{L}_{adv}(\mathcal{T},\theta^K)\cdot \nabla_{\theta^t}r'_{\theta^t}(\hat{\mathcal{G}})\cdot \nabla_{\hat{\mathbf{R}}}\theta^K,
\end{aligned}
\end{equation}
where $\nabla_{\hat{\mathbf{R}}}\theta^{t+1}=\nabla_{\hat{\mathbf{R}}}\theta^{t}-\eta_1\nabla_{\hat{\mathbf{R}}}\nabla_{\theta^t}\mathcal{L}(\theta, \hat{\mathcal{G}})$, $t=0,\cdots, K-1$, $\eta_1$ is the learning rate of the inner model, $r'_{\theta^t}(\hat{\mathcal{G}})=\{U_u^TV_v|\forall (u,v)\in \hat{\mathcal{G}}\}$, are the rating predictions between among user-item pairs of the inner model.

The difficulty, however, lies in that it requires the loss function $\mathcal{L}(\theta, \hat{\mathcal{G}})$ in Eqn~\eqref{loss-for-tensor} is differentiable with respect to $\hat{\textbf{R}}$. To address this, we smooth the $\argmax$ function  via $softmax$ function. 
We approximate the $\argmax$ function by $softmax$ with parameter $\beta$ (the larger $\beta$, the better approximation) as follow,
\begin{equation}
   \argmax_l \hat{\mathbf{R}}_{uvl} \approx [1,2,3,4,5]^T \times softmax(\beta \cdot \hat{\mathbf{R}}_{uv}).
\end{equation}
Note that the attack optimized on the MF-based model is only different in the calculation of $\nabla_{\hat{\mathbf{R}}}^{meta} \mathcal{L}_{adv}$, and the other processes are the same.

\subsection{\textbf{PDR}}
To adapt our approach to MF-based RS, we substitute the loss function Eqn~\eqref{PDR-loss-function} with the following:
\begin{equation}
\setlength{\jot}{0.01pt}
\begin{aligned}
\mathcal{L}_{\text{rating}} &={\textstyle \frac{1}{\left| \mathcal{E} \right| }} \sum_{\forall (u,v) \in \mathcal{E}} q_{\theta}(l_u=n|z_u) \cdot (r_{uv}-U_u^TV_v)^2.\nonumber
\end{aligned}
\end{equation}
Without the GNNs module that can provide user embedding, we estimate the posterior probability using a single-layer GNN and an MLP:
$q(l_u|z_u)=softmax_T(\text{MLP}(z_u))$, where $z_u=GNN(\mathcal{G}')$, the input is poisoned history rating graph $\mathcal{G}'$. Specifically, we normalize the ratings as the weights on edges, and employ single layer GraphSAGE~\cite{hamilton2017inductive} with mean aggregator:
\begin{equation}
    z_u=ReLU (W \cdot ( \sum_{v\in \mathcal{N}(u)} \frac{h_v \cdot \omega_{u,v}}{\sum_{i \in \mathcal{N}(u)}\omega_{u,i}} \oplus h_u),
\end{equation}
where $W$ is learnable weight matrix, and $h_v$ is initial embedding of user or item, $ReLU(\cdot)$ is activation function, $\oplus$ is concatenating function, $\omega_{u,v} \in [0,1]$ is the normalized rating between user $u$ and item $v$.

\section{Experiments}
\label{Experiments}
In section, we aim to evaluate our methods by answering the following key questions:
\begin{itemize}
\item Does the \textbf{MetaC} attack effectively compromise the security of existing robust recommendation systems (\ref{Question1})?
\item How resilient is our proposed \textbf{PDR} framework against  attacks (\ref{Question2})?
\item What are the underlying mechanisms that enable \textbf{PDR} to achieve adversarial robustness (\ref{Question3})?
\item How does the level of prior knowledge $\tau$ impact the performance of \textbf{PDR} (\ref{Question4})?
\end{itemize}

\subsection{Datasets and Experiment Settings}
\paragraph{Datasets} 
We conduct experiments over two widely-used real-world datasets \textit{YelpCHI} and \textit{Amazon Movies\&TV} (abbreviated as \textit{Movies}) that collect user reviews from two platforms.
\textit{YelpCHI} contains approximately $60,000$ reviews/ratings regarding $201$ restaurants and hotels in Chicago from $38,063$ reviewers. Each rating ranges from $1$ to $5$, and the corresponding review is provided with a label of \textit{fake} or \textit{normal}. In our setting, we treat a user giving \textit{fake} review(s) as \textit{fake}. The other dataset \textit{Movies} contains reviews from Amazon under the category of Movie\&TV. Each review, with a rating from $1$ to $5$, is voted \textit{helpful/unhelpful} by other users, which provides the information to determine whether a user is fake or normal. Specifically, we only consider the reviews with more than 20 votes. If more than $70\%$ of the votes of a review are \textit{helpful}, we regard it as \textit{normal}; otherwise, \textit{fake}. Similarly to \textit{YelpCHI}, we treat the users giving \textit{fake} review(s) as \textit{fake} users. The statistics of the two datasets are summarized in Tab.~\ref{table-dataset}.
\begin{table}[h]
\caption{Statistics of \textit{YelpCHI} and \textit{Movies}.}
\footnotesize
\label{table-dataset}
\centering
\begin{tabular}{lrrrr}
	\toprule[0.8pt]
		& \# Users & \# Items & \# Edges & \# Fake users \\ \hline
		\textit{YelpCHI}  & $38063$         & $201$        & $67395$        & $7739$              \\
		\textit{Movies} & $39578$   & $71187$   & $232082$  & $19909$        \\ 
         \toprule[0.8pt]
	\end{tabular}
\end{table}

Note that there are two types of fake users, one is the existing fake users in the dataset (we do not know their targets/goals, but these users can enlarge the number of anomaly users for detection model training), and the other is our own injected fake user (from 0.3\% to 2.0\% according to attack power setting). During the data preprocessing, we iteratively remove the cold items and users that are less than two records. We extract the user features used in GNNs following \cite{zhang2020gcn}. 

\paragraph{Environments} We conduct our experiments on Intel 10C20T Core i9-10850K CPU with GIGABYTE RTX3090 24GB GPU on development environment Ubuntu18.04, Python 3.7, PyTorch 1.10.0. 

\paragraph{Settings} 
Following the typical settings in \cite{zhang2021data}, we randomly sample $5$ items from all items as the targets. To train the RS model, we randomly sample $20\%$ of existing ratings labeled with normal for testing, and the remaining are the training set. The rating budget of each injected user is $B=15$. Due to the large number of items in the \textit{Movies} dataset, the search space for the attack optimization is extensive. As noted in \cite{takahashi2019indirect}, two-layer GCN is highly susceptible to poisoning nodes within 2-hop. Therefore, we limit the space of candidate items to 2-hop neighbors of the target items, which improves the efficiency of the training process. The experiments are repeated for $5$ times with different random seeds to initialize model parameters. 
We use the averaged hit ratios of the target items, i.e., $HR@10$ and $HR@50$, as the metrics to evaluate how the attacks can promote target items. We test the attack performances under various attack powers ($0.0\%$, $0.3\%$, $0.5\%$, $0.7\%$, $1.0\%$, $2.0\%$), 
where an attack power represents the fraction of the number of injected users over all users. In this section, the fraction of injected fake users with correct labels is set as $\tau = 30\%$ to reflect that a user deploying \textit{GraphRfi} may have some prior knowledge about the data. However, later we show that \textit{GraphRfi} can be successfully attacked regardless of the value of $\tau$. 
In attacks, the number of inject user proportion is set to 1\% of the original user number and detection fraction $\tau = 30\%$ if not mentioned. We set the alternate iteration steps $K=100$, batch size 128. We totally train the model for $T_{train}$ epochs ($T_{train}=50$ for \textit{GraphRfi}; $T_{train}=300$ for MF-based model) during attack optimization and retrain the model after poisoning for $T_{train}$ epochs. The smoothing parameter $\beta$ for attacking the MF-based model is set to $10.0$.
In the PDR framework, we set the temperature of $softmax$ function as $T=2.0$. We set the probability that a user labeled with fake is actually normal as $p_0=0.01$, and the probability that a user labeled with normal is actually fake as $p_1=0.2$. In the label adjustment strategy, we update the labels when the AUC of the detection model reaches $a_0$ ($a_0=0.8$ for \textit{GraphRfi}; $a_0=0.7$ for MF-based) on the training set, and we set update rate $\alpha=0.05$ to adjust priors. At the beginning, we set the adjusting interval parameters $c_1=0.4$, $c_2=0.85$, and decreasing $c_1$ while increasing $c_2$ to decay the range of adjusting interval by $c_1^{t+1}=\min \{c_1^{t}-0.025,0.2\}$, and $c_2^{t+1}=\max \{c_2^{t}+0.025,1.0\}$. We set a larger adjust interval for the normal user side ($q(f_u)^{t} < c_1$) since there are more normal users than fake users. Embedding dimension in \textit{GraphRfi} is set as $50$ for \textit{YelpCHI} and $100$ for \textit{Movies}; $128$ in MF. The hidden layer number of MLP is $2$. The regularization coefficient is set as $0.01$ in \textit{GraphRfi} and $1\times10^{-5}$ in \textit{MF}.

\subsection{Baseline Methods}
\paragraph{Attack} We compare to five representative attack methods: 
\textit{Random}, \textit{Average}, \textit{Popular/Bandwagon} \cite{wu2021ready}, \textit{Poison Training} (\textit{PoisonT}) \cite{Hai2021data}, and \textit{Trial Attack} \cite{wu2021triple}. Among these attacks, \textit{Random}, \textit{Average}, and \textit{Popular/Bandwagon} do not depend on the RS model while \textit{PoisonT} and \textit{Trial Attack} are designed for MF-based RS. Each fake user gives the highest ratings to the target items and rates a set of \textit{filler} items using the remaining budget with various strategies. In \textit{Random Attack}, filler items are randomly selected, and the corresponding ratings are sampled from a normal distribution. 
$\mathcal{N}(\mu,\sigma^2)$, where $\mu$ and $\sigma$ are the mean and deviation of all existing ratings. For \textit{Average Attack}, the only difference from \textit{Random Attack} is that the rating given to a filler item $v_i$ is sampled from $\mathcal{N}(\mu_{v_{i}},\sigma_{v_{i}})$, where the $\mu_{v_{i}},\sigma_{v_{i}}$ are the means and deviation of existing ratings for item $v_i$. In \textit{Popular Attack}, a portion (set as $30\%$ in our experiment) of filler items are selected as popular items since they might have bigger impacts, and the ratings given to these popular items are also set as $r_{max}$. In \textit{PoisonT Attack}, it adds poisoning users one by one with maximum ratings given to target items and trains a poisoned RS model to predict the ratings for filler items. It chooses filler items in descending order of the predicted rating scores and gives ratings sampled from a normal distribution. \textit{Trial Attack} trains generator module, influence module, and discrimination module together to generate stealthy fake users that maximise the influence on attack goals while evading the detection of the discriminator. 

\paragraph{Defense} Since there are no prior defense strategies for \textit{GraphRfi}, we adapt representative defense approaches from both categories. First, for the adversarial training approach, 
we adopt a representative work proposed by ~\cite{yuan2019adversarial}. Specifically, it adds \textit{perturbation noise} to the model parameters when training the model. 
Second, we explore the idea of using the result of anomaly detection for defense. The natural idea is to remove the detected fake users from the system. Note that, we do not constrain a specific method here for anomaly detection; instead, we assume that a fraction $\tau$ of the injected fake users can be detected due to the fact that any anomaly detection method might be employed in practice and the detection performance varies. In our experiment, this fraction $\tau$ of fake users are removed; we thus term this approach as \textit{Remove Anomaly}.

\subsection{Effectiveness of \textbf{MetaC} Attack}
\label{Question1}
\paragraph{Againt MF} We start with MF-based RS, for which there exists a direct comparison. The attack results are summarized in Tab.~\ref{attackMF_YelpCHI} and Tab.~\ref{tab:attack-MF}. We can observe that \textbf{MetaC} achieves the best results under all attack powers, demonstrating its strength as well as the challenge in defending against this strong attack. Besides, the attack performance of \textbf{MetaC} is still better than \textit{PoisonT} most of the time under the two defense frameworks with anomaly detection (highlighted by underlining in Tab.~\ref{tab:defense-MF-YelpCHI} and Tab.~\ref{tab:defense-MF-Movies}), which further demonstrates the power of \textbf{MetaC}. 
In addition, \textit{PoisonT} has slightly better attack performance than that of \textit{Trial Attack}. Also, due to its computational simplicity, we choose to adapt \textit{Poison Training} for attacking \textit{GraphRfi}.

\begin{table}[h]
\caption{Attack results ($HR@50$) on MF-based model (\textit{YelpCHI}).}
\label{attackMF_YelpCHI}
\footnotesize
\setlength{\tabcolsep}{4pt}
\centering
\begin{tabular}{lrrrrrr}
\toprule[0.8pt]
Power  & \textit{Random} & \textit{Average} & \textit{Popular} & \textit{PoisonT} & \textit{Trial} & \textbf{MetaC}          \\ \toprule[0.8pt]
0.0\% & 0.389  & 0.389   & 0.389   & 0.389   & 0.389  & 0.389          \\ \hline
0.3\% & 0.446  & 0.445   & 0.462   & 0.471   & 0.551  & \textbf{0.614} \\
0.5\% & 0.529  & 0.524   & 0.500   & 0.523   & 0.638  & \textbf{0.772} \\
0.7\% & 0.584  & 0.587   & 0.540   & 0.661   & 0.694  & \textbf{0.864} \\
1.0\% & 0.682  & 0.673   & 0.596   & 0.776   & 0.717  & \textbf{0.929} \\
2.0\% & 0.909  & 0.889   & 0.773   & 0.942   & 0.790  & \textbf{0.972} \\ 
\toprule[0.8pt]
\end{tabular}
\end{table}

\begin{table}[h]
\caption{Attack results ($HR@50$) on MF (\textit{Movies}).}
\label{tab:attack-MF}
\footnotesize
\setlength{\tabcolsep}{4pt}
\centering
\begin{tabular}{lrrrrrr}
\toprule[0.8pt]
Power &\textit{Random} & \textit{Average} & \textit{Popular} & \textit{PoisonT} & \textit{Trial} & \textbf{MetaC} \\ \toprule[0.8pt]
0.0\% & 0.200 & 0.200 & 0.200 & 0.200 & 0.200 & 0.200 \\ \hline
0.3\% & 0.376 & 0.371 & 0.258 & 0.386 & 0.398 & \textbf{0.409} \\
0.5\% & 0.387 & 0.377 & 0.398 & 0.397 & 0.408 & \textbf{0.522} \\
0.7\% & 0.398 & 0.392 & 0.384 & 0.401 & 0.409 & \textbf{0.828} \\
1.0\% & 0.418 & 0.399 & 0.404 & 0.445 & 0.507 & \textbf{0.957} \\
2.0\% & 0.787 & 0.550 & 0.534 & 0.929 & 0.620 & \textbf{0.986} \\ 
\toprule[0.8pt]
\end{tabular}
\end{table}

\begin{figure}[t]
\centering
    \subfigure[YelpCHI]{
    \includegraphics[width=0.225\textwidth,height=3.0cm]{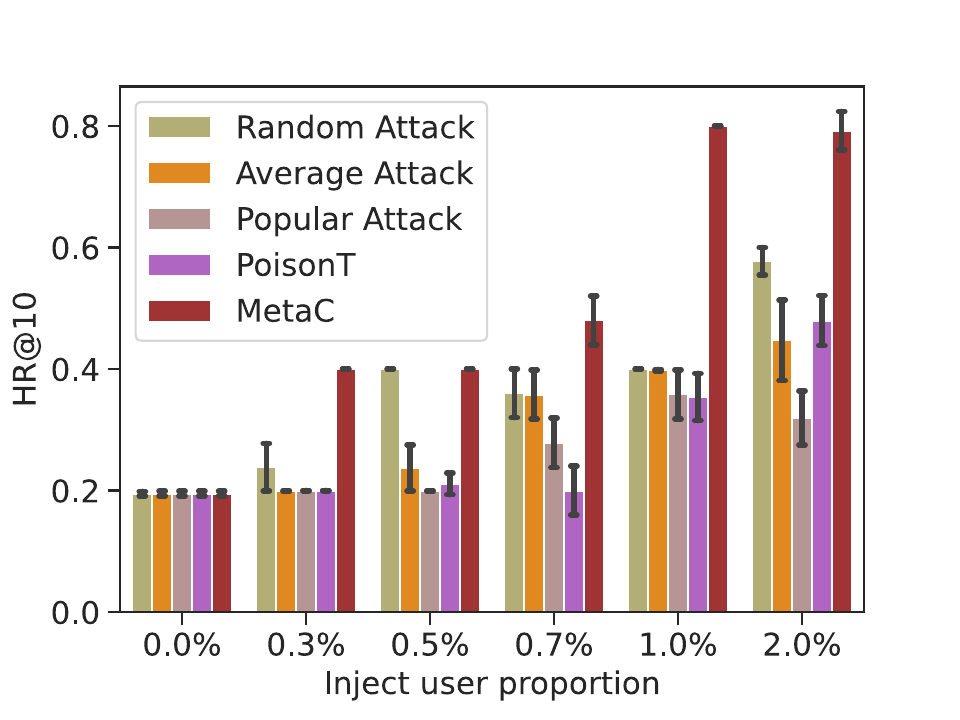}
    }
    \subfigure[YelpCHI]{
    \includegraphics[width=0.225\textwidth,height=3.0cm]{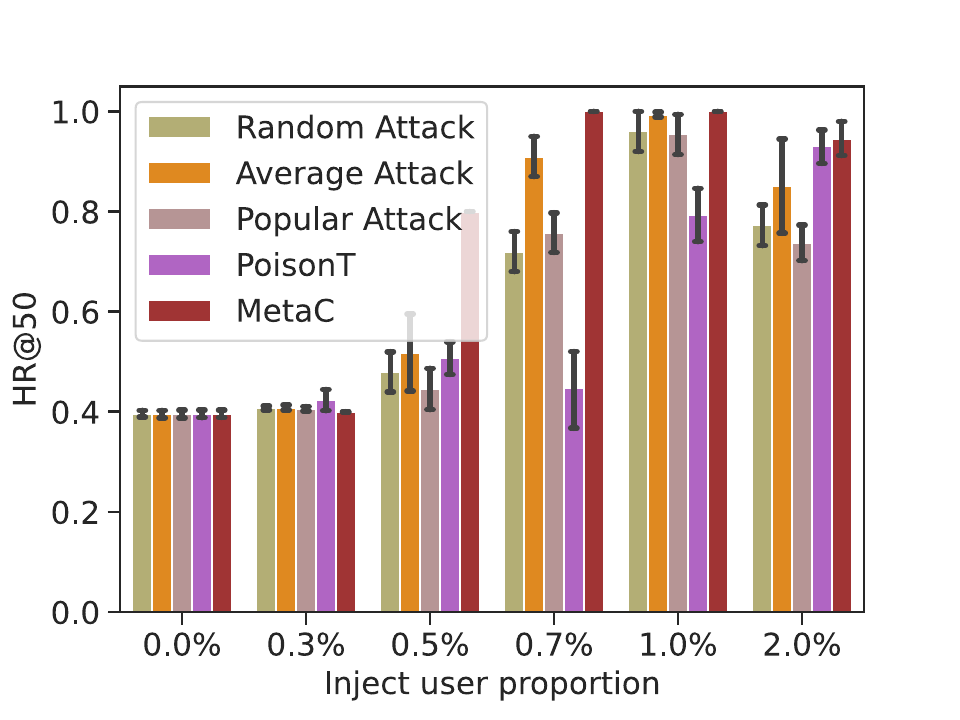}
    }
    \subfigure[Movies]{
    \includegraphics[width=0.225\textwidth,height=3.0cm]{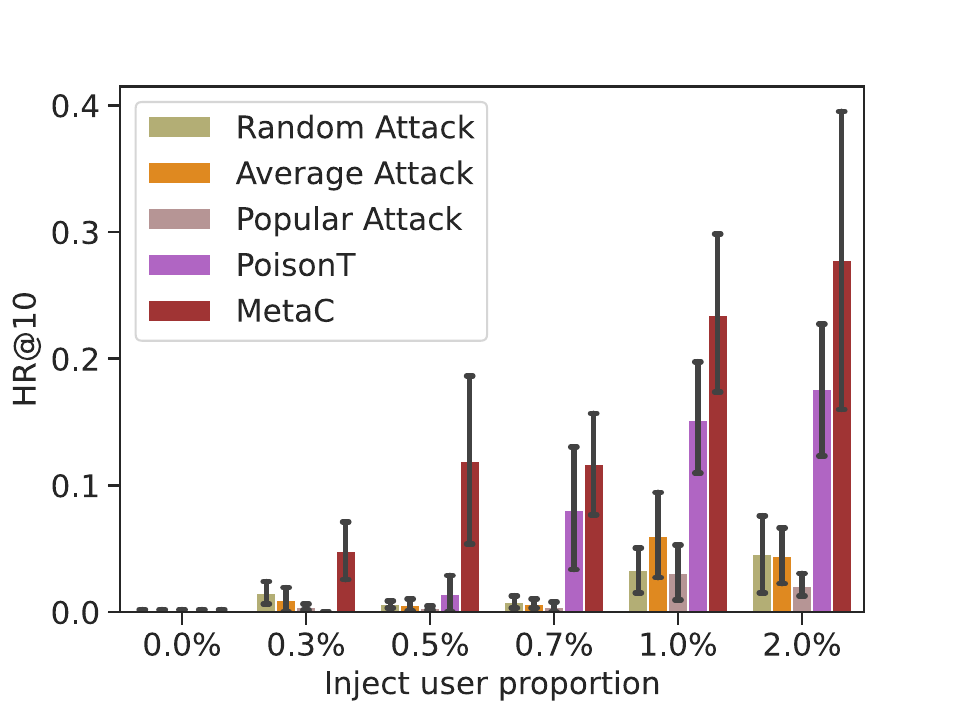}
    }
    \subfigure[Movies]{
    \includegraphics[width=0.225\textwidth,height=3.0cm]{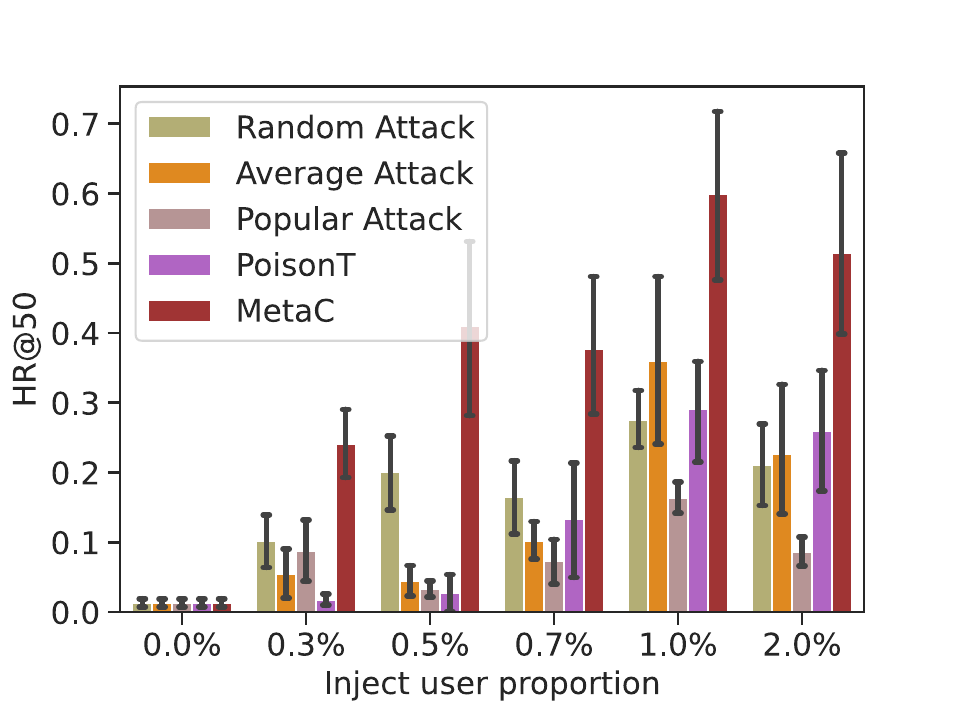}
    }
\caption{Attack performances with different powers}
\label{fig:attack}
\end{figure}

\paragraph{Against \textit{GraphRfi}}
The hit ratios under attack (the higher, the better) are presented in Fig.~\ref{fig:attack}. We can see that \textbf{MetaC} achieves the best attack performances in all cases, especially on \textit{Movies}. One observation is that the gaps between \textbf{MetaC} and others are more evident for $HR@10$. Note that pushing the items to top-10 is harder than pushing to top-50, which further demonstrates the effectiveness of \textbf{MetaC}. Meanwhile, we notice that the results show a larger variance on \textit{Movies}. A possible reason is that hit ratio is a ranking-based metric, and \textit{Movies} has significantly more items. Thus, there are much more items around the top-10/50 threshold, making the ranking sensitive to perturbations (this large variance is also observed in related research \cite{tang2020revisiting,oh2022rank}).

\subsection{Robustness Evaluation of \textbf{PDR}}
\label{Question2}
\begin{figure}[!ht]
	\centering
	\subfigure[YelpCHI]{
	\includegraphics[scale = 0.28]{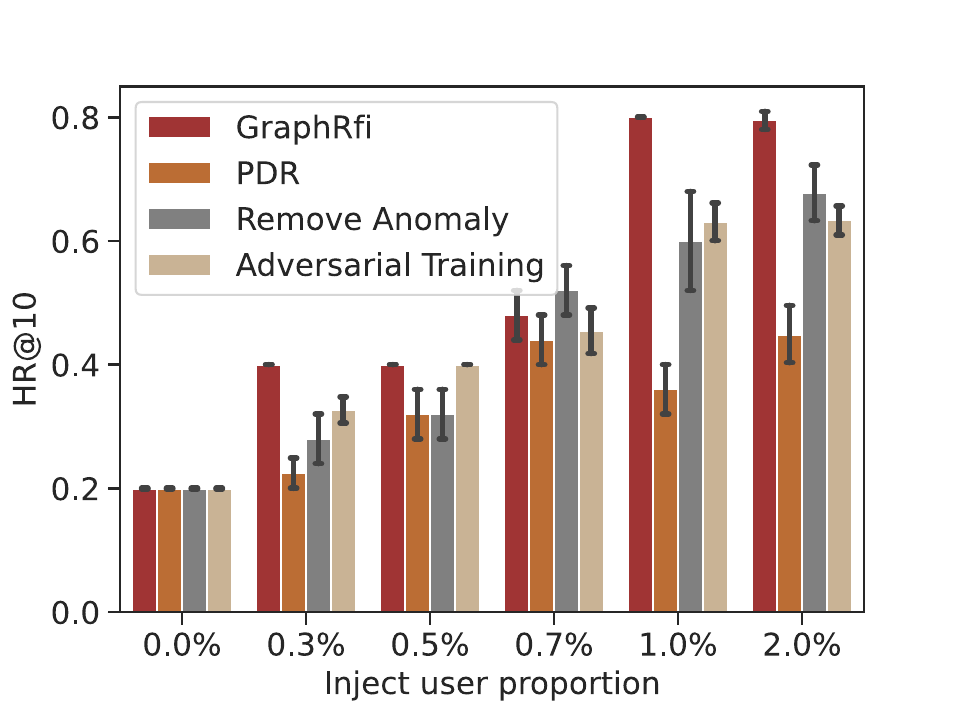}
	}
	\subfigure[YelpCHI]{
	\includegraphics[scale = 0.28]{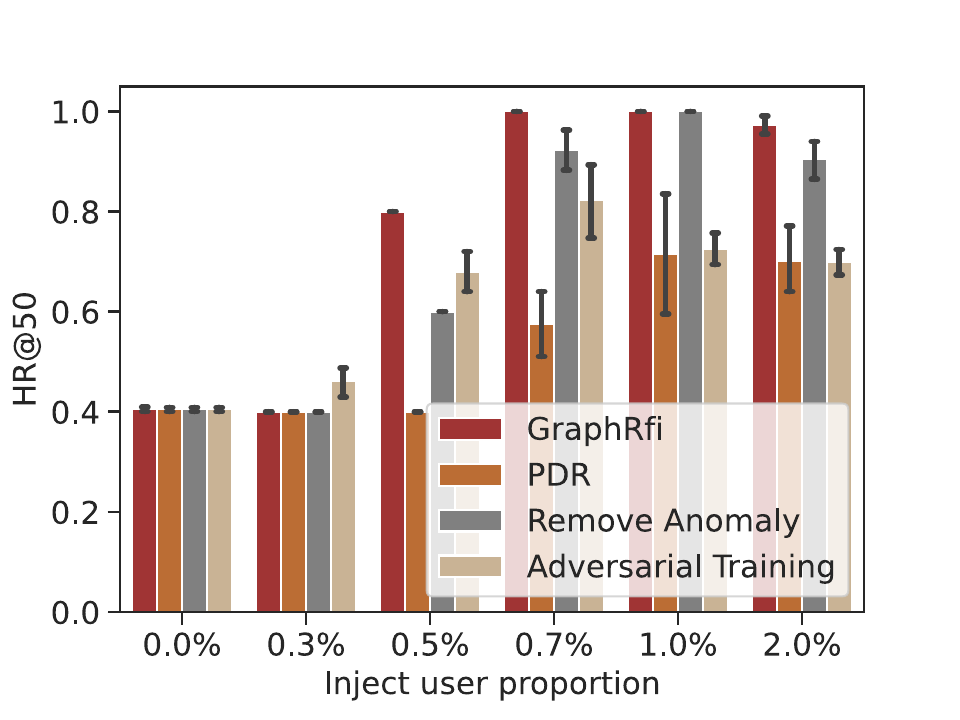}
	}
	\subfigure[Movies]{
	\includegraphics[scale = 0.28]{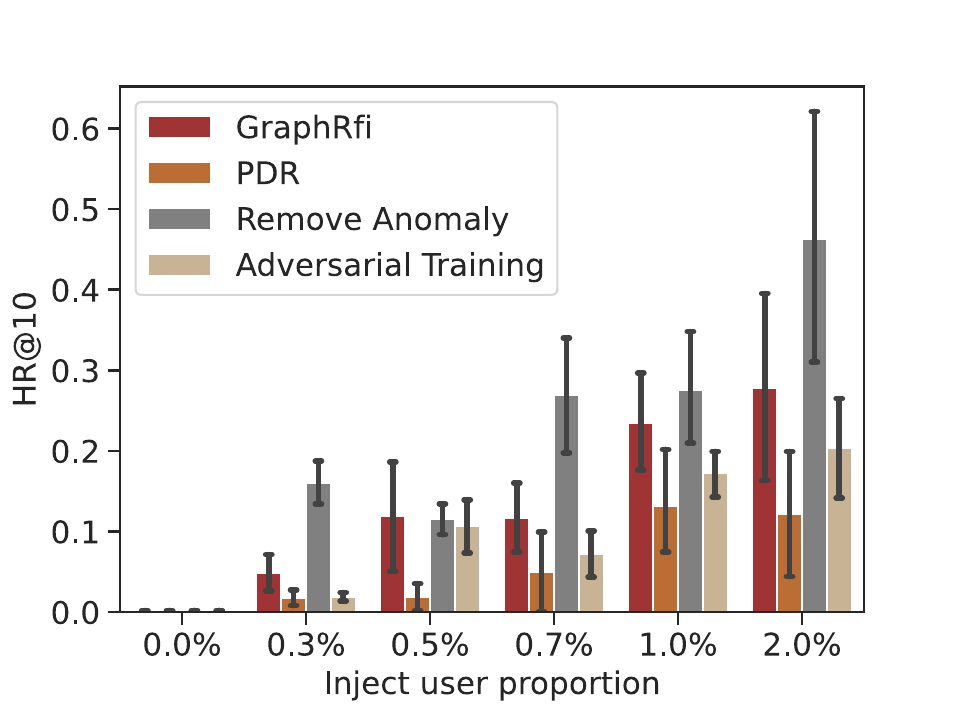}
	}
	\subfigure[Movies]{
	\includegraphics[scale = 0.28]{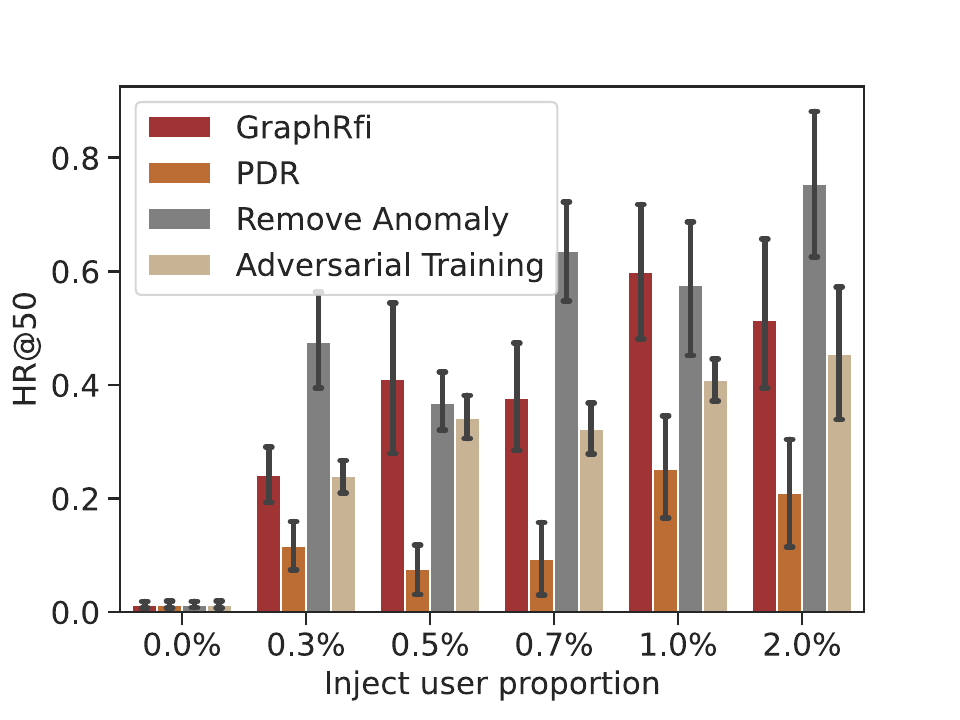}
	}
        \caption{Defense performances under different powers.}
	\label{fig:defense}
\end{figure}

The primary defense goal is to retain the hit ratios of the target items under attack. Fig.~\ref{fig:defense} presents the performances of different defense approaches applied to \textit{GraphRfi}. We can see that \textbf{PDR} achieves the best defense performance, especially when the attack power is higher (it is also when defense is harder).  
We note that \textit{Remove Anomaly} may or may not be better than \textit{GraphRfi} (i.e., without defense). The reason is that the anomaly detection component within \textit{GraphRfi} is supervised. Thus, removing the correctly labeled fake users, as \textit{Remove Anomaly} did, reduces the supervision, which might harm the performance. This actually demonstrates the significance of our proposed way of dealing with those detected fake users. Similarly, the advantages of \textbf{PDR} on MF-based RS can also be observed. Tab.~\ref{tab:defense-MF-Movies} and Tab.~\ref{tab:defense-MF-YelpCHI} shows the defense performance on MF-based model under $2$ attacks (\textit{PoisonT} and \textbf{MetaC}), where \textit{No defense} is the original MF-based model, \textit{Hard label} adds the same GNN detection model in \textbf{PDR}, but it uses the common cross-entropy loss with hard labels (similar to \textit{GraphRfi}), and \textbf{PDR} is our robust model that uses soft label posterior detection. The results demonstrate that the MF-based model trained with \textbf{PDR} has the closest $HR@10$ to the original one with $0\%$ attack power. 

\begin{table}[t]
\footnotesize
\centering
\setlength{\tabcolsep}{1.5pt}
\caption{Defense performance ($HR@10$) on MF-based model (\textit{YelpCHI}).}
\label{tab:defense-MF-YelpCHI}
\begin{tabular}{c|ccc|ccc}
\toprule[0.8pt]
Attack Method & \multicolumn{3}{c|}{\textit{PoisonT}}                & \multicolumn{3}{c}{\textbf{MetaC}}                \\ \toprule[0.8pt]
Attack Power          & No defense & Hard label & \textbf{PDR}            & No defense & Hard label & \textbf{PDR}            \\ \toprule[0.8pt]
$0.0\%$         & 0.214      & 0.214      & 0.214          & 0.214      & 0.214      & 0.214          \\ \hline
$0.3\%$         & 0.245      & 0.235      & \textbf{0.227} & 0.335      & \underline{0.261}      & \underline{\textbf{0.240}} \\
$0.5\%$         & 0.265      & 0.243      & \textbf{0.238} & 0.426      & \underline{0.281}      & \underline{\textbf{0.263}} \\
$0.7\%$         & 0.279      & 0.241      & \textbf{0.231} & 0.537      & \underline{0.299}      & \underline{\textbf{0.273}} \\
$1.0\%$         & 0.302      & 0.265      & \textbf{0.252} & 0.701      & \underline{0.339}      & \underline{\textbf{0.309}} \\
$2.0\%$         & 0.467      & 0.304      & \textbf{0.267} & 0.880      & \underline{0.401}      & \underline{\textbf{0.364}} \\ 
\toprule[0.8pt]
\end{tabular}
\end{table}

\begin{table}[t]
\footnotesize
\centering
\setlength{\tabcolsep}{1.5pt}
\caption{Defense performance ($HR@10$) on MF-based model (\textit{Movies}).}
\label{tab:defense-MF-Movies}
\begin{tabular}{c|ccc|ccc}
\toprule[0.8pt]
Attack Method & \multicolumn{3}{c|}{\textit{PoisonT}}                & \multicolumn{3}{c}{\textbf{MetaC}}                \\ \toprule[0.8pt]
Attack Power          & No defense & Hard label & \textbf{PDR}            & No defense & Hard label & \textbf{PDR}            \\ \toprule[0.8pt]
$0.0\%$         & 0.183      & 0.183      & 0.183          & 0.183      & 0.183      & 0.183          \\ \hline
$0.3\%$         & 0.201      & \underline{0.274}      & \underline{\textbf{0.200}} & 0.350      & 0.264      & \textbf{0.199} \\
$0.5\%$         & 0.302      & 0.271      & \textbf{0.199} & 0.371      & \underline{0.334}      & \underline{\textbf{0.322}} \\
$0.7\%$         & 0.359      & 0.272      & \textbf{0.199} & 0.424      & \underline{0.456}      & \underline{\textbf{0.362}} \\
$1.0\%$         & 0.352      & 0.282      & \textbf{0.200} & 0.606      & \underline{0.563}      & \underline{\textbf{0.539}} \\
$2.0\%$         & 0.407      & 0.382      & \textbf{0.220} & 0.923      & \underline{0.624}      & \underline{\textbf{0.557}} \\ 
\toprule[0.8pt]
\end{tabular}
\end{table}



\subsection{Why \textbf{PDR} is robust}
\label{Question3}
The adversarial robustness of \textbf{PDR} comes from the fact that it can detect and dynamically adjust the contributions of fake users in the recommender system.

\begin{figure}[t]
	\centering
	\subfigure[without defense (\textit{YelpCHI})]{
    \includegraphics[width=0.225\textwidth,height=3.0cm]{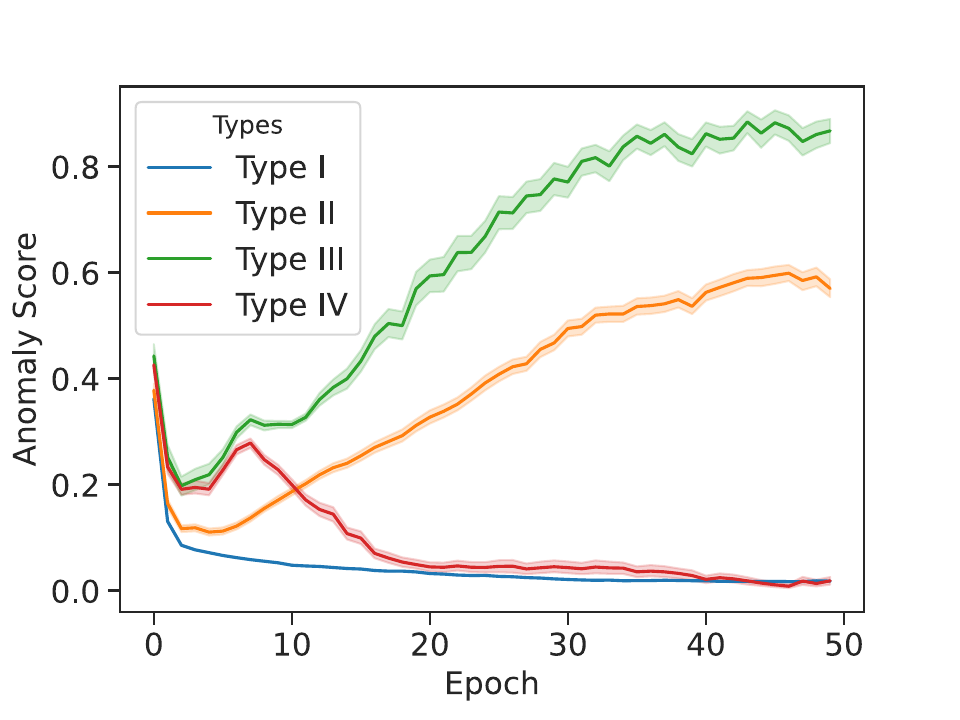}
		\label{fig-score-no-defense-yelp}
	}
	\subfigure[with defense (\textit{YelpCHI})]{
		\includegraphics[width=0.225\textwidth,height=3.0cm]{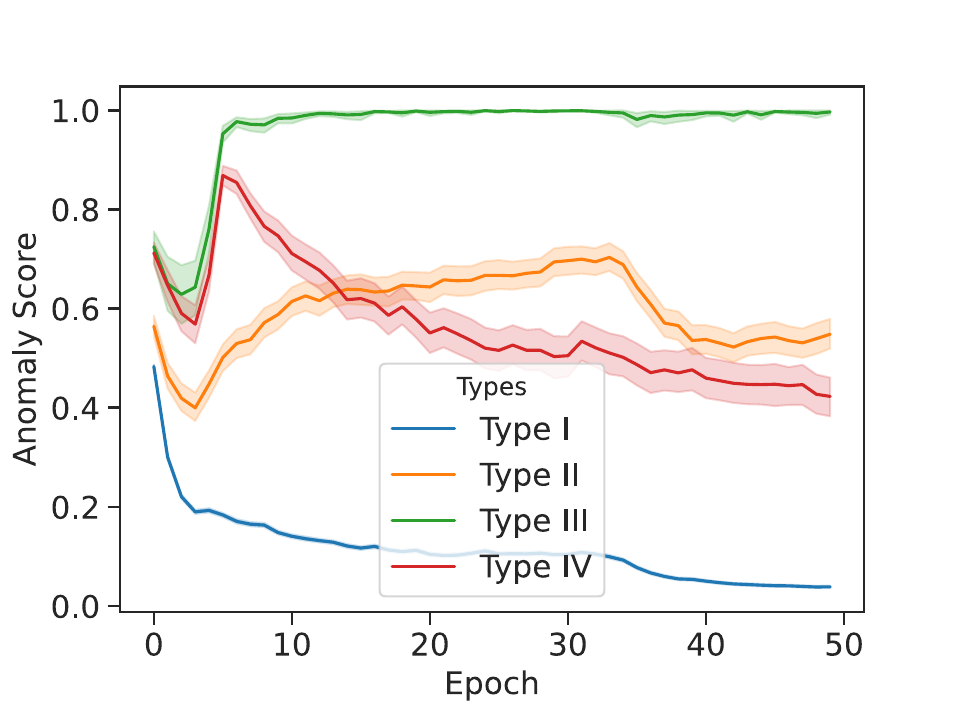}
		\label{fig-score-defense-yelp}
	}
         \subfigure[without defense (\textit{Movies})]{
		\includegraphics[width=0.225\textwidth,height=3.0cm]{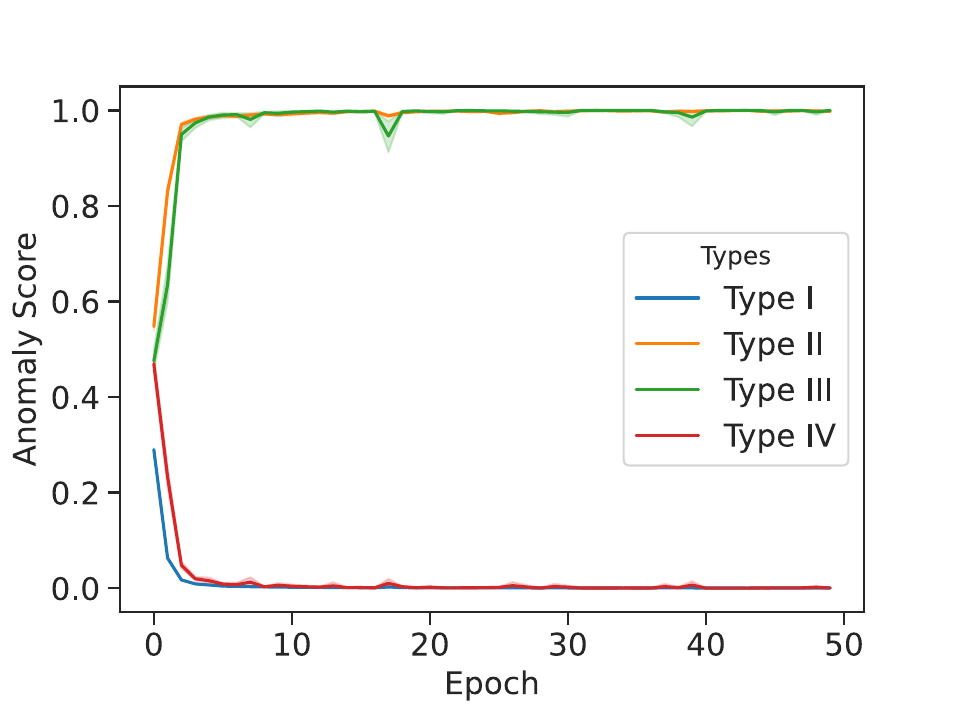}
	}
	\subfigure[with defense (\textit{Movies})]{
		\includegraphics[width=0.225\textwidth,height=3.0cm]{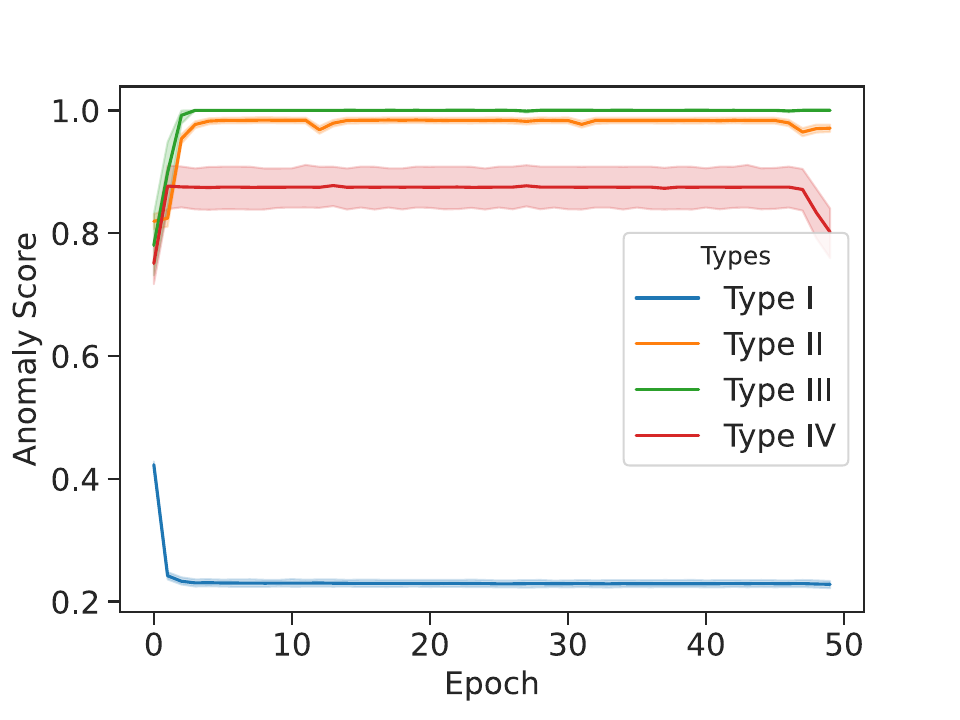}
	}
\caption{Anomaly scores for different types of users.}

\end{figure}

To illustrate this point, we visualize the trajectories of anomaly scores (inversely proportional to contribution) of different types of users during the training of two systems \textit{GraphRfi} and \textbf{PDR} in Fig.~\ref{fig-score-no-defense-yelp} and Fig.~\ref{fig-score-defense-yelp}, respectively. What we should focus on is the Type IV users (i.e., fake users but labeled as normal), the anomaly scores of which are shown in red over \textit{YelpCHI}. Compared to \textit{GraphRfi}, \textbf{PDR} can assign large anomaly scores even for Type IV users, which is the reason for its adversarial robustness. The results over \textit{Movies} are similar.

\subsection{Influence of Prior Knowledge}
\label{Question4}
\begin{figure}[h]
	\centering
	\subfigure[YelpCHI]{
		\includegraphics[width=0.225\textwidth,height=3.0cm]{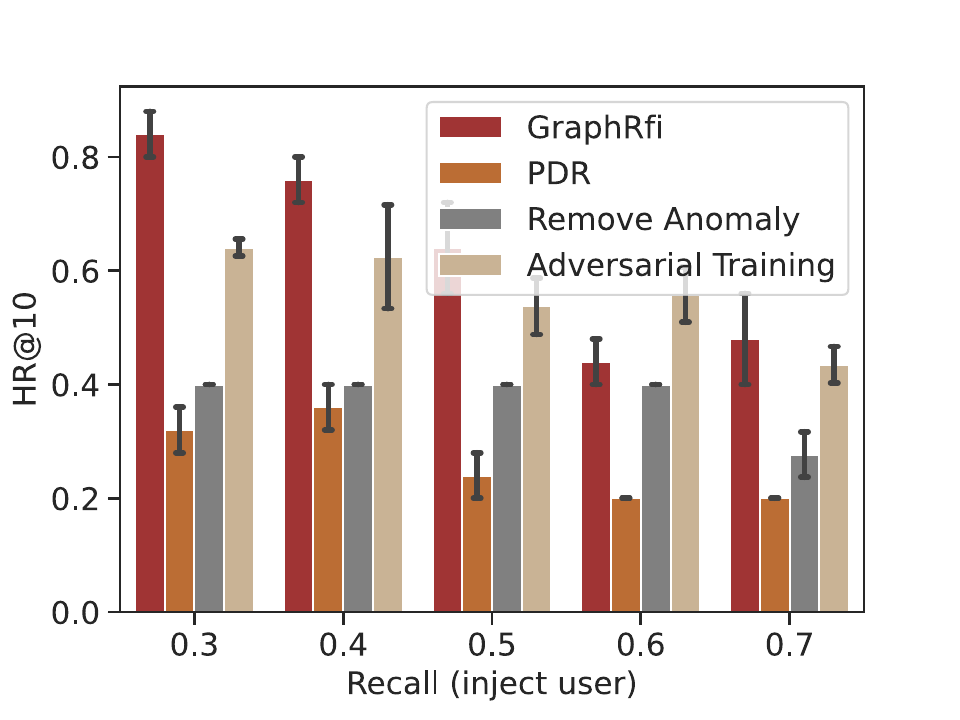}
	}
	\subfigure[YelpCHI]{
		\includegraphics[width=0.225\textwidth,height=3.0cm]{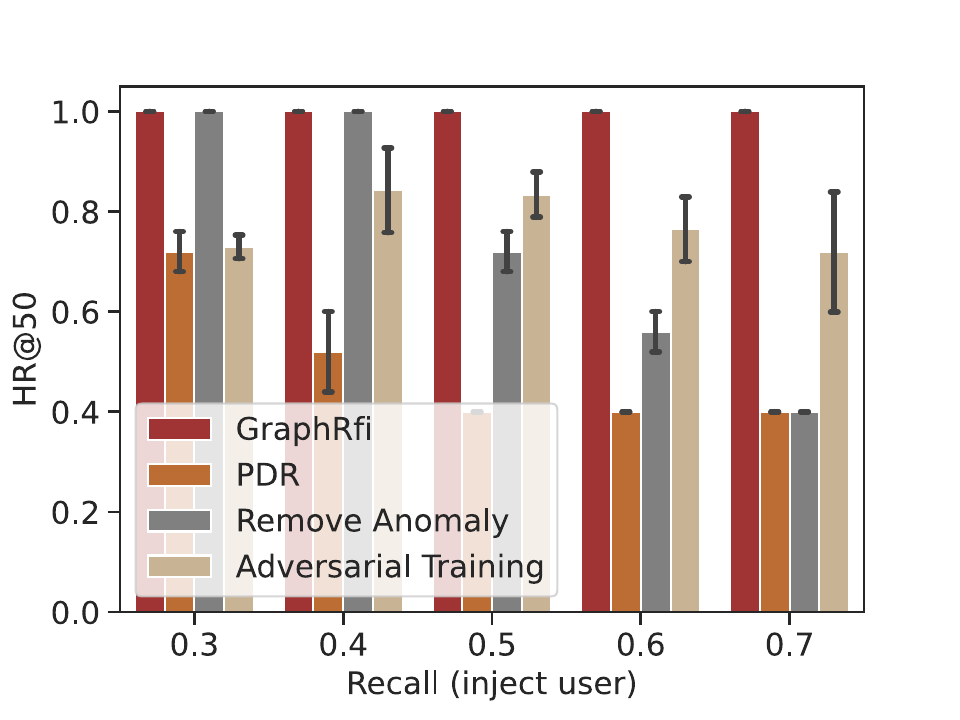}
	}
  	\subfigure[Movies]{
		\includegraphics[width=0.225\textwidth,height=3.0cm]{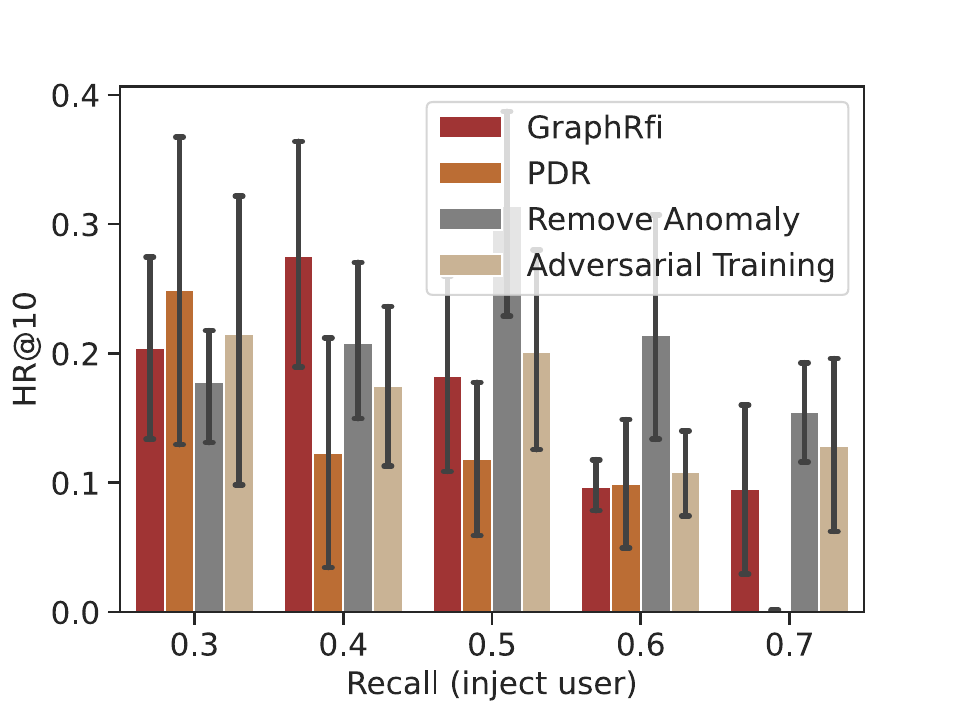}
	}
	\subfigure[Movies]{
		\includegraphics[width=0.225\textwidth,height=3.0cm]{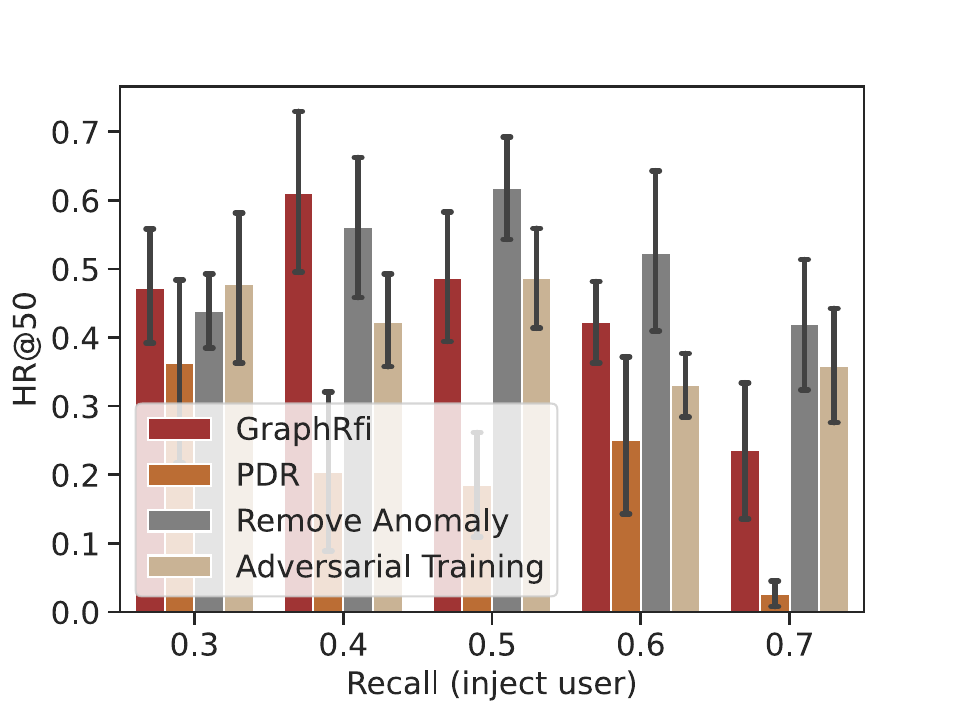}
	}
	\caption{Defense performance with various $\tau$ on \textit{YelpCHI}.}
	\label{fig:recall}
\end{figure}

In the experiments, we use a parameter $\tau$ to control the defender's prior knowledge (possibly obtained from using some anomaly detection methods to preprocess the data) regarding the injected fake users. Specifically, $\tau$ is the \textit{recall} over injected users defined as $\tau= \frac{|\{u\in\mathcal{U}'| \text{labeled as fake}\}|}{|\mathcal{U}'|}$, representing the fraction of fake users that are correctly labeled. We thus evaluate the two different ways (i.e., \textit{Remove Anomaly} and \textbf{PDR}) of dealing with detected fake users under different levels of $\tau$. Fig. \ref{fig:recall} shows that \textbf{PDR} achieves the best performance over \textit{YelpCHI} and the performance becomes better as $\tau$ increases as it receives more supervision. Again, \textit{Remove Anomaly} is not quite effective in some cases as removing correctly labeled fake users also decreases the supervision.

\subsection{Hyper-parameter}
\begin{figure}[htbp]
	\centering
	\subfigure[]{
		\includegraphics[width=0.225\textwidth,height=3.0cm]{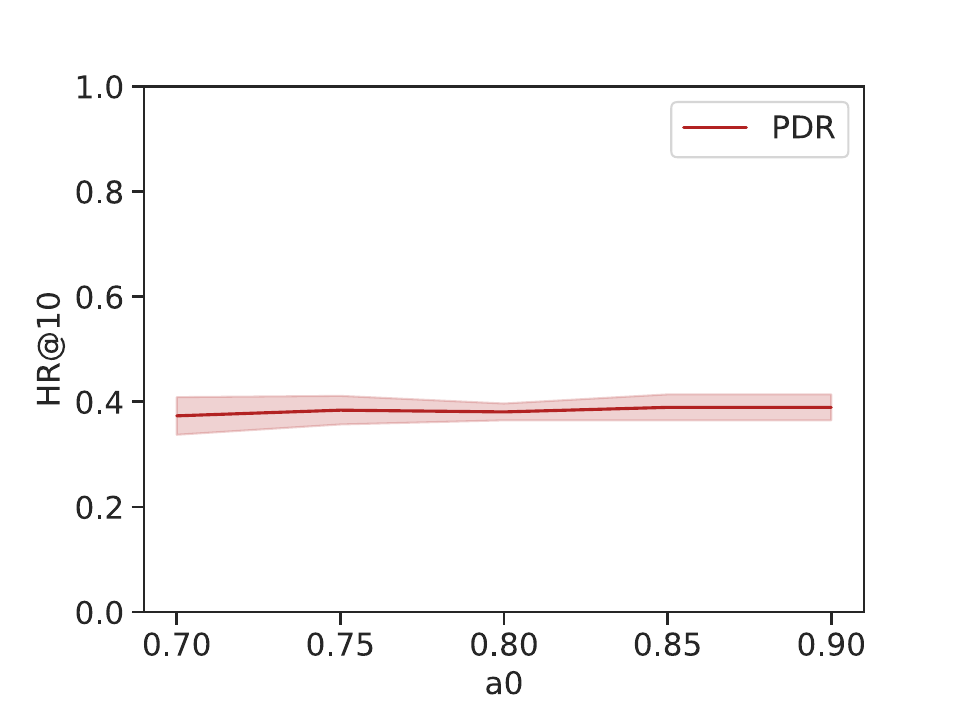}
	}
	\subfigure[]{
		\includegraphics[width=0.225\textwidth,height=3.0cm]{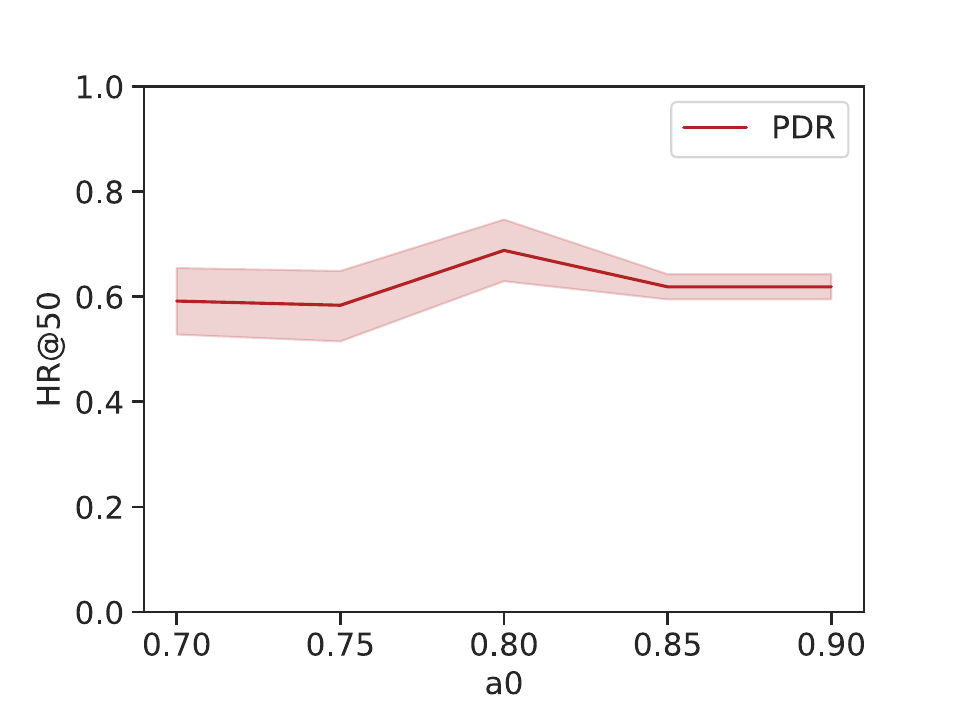}
	}
	
	\caption{Hyparameter $a_0$: HR@10/50 of target items, under different hyparameter $a_0$ on \textit{YelpCHI} dataset, with inject user proportion 0.1\%.}
	\label{fig:a_0}
\end{figure}

To evaluate the sensitivity of hyper-parameter $a_0$ which decides when the label adjustment begins, we conduct experiments based on $a_0$ from 0.7 to 0.9 on the dataset \textit{YelpCHI}. And the results in Fig. \ref{fig:a_0} show that, the defense effectiveness is robust under different $a_0$. 



\section{Conclusion}
\label{Conclusion}
In this paper, we demonstrated the vulnerabilities of a state-of-the-art robust recommender system called \textit{GraphRfi} by designing an effective attack approach \textbf{MetaC}. 
We re-designed the detection component which is equipped with the ability to dynamically adjust the importance of \textit{newly} injected fake users, resulting in a robust RS termed \textbf{PDR}. In addition, we also show that our attack and defense methods can also be applied to MF-based RS.
This research demonstrated the effectiveness of a framework for integrating anomaly detection into learning systems to improve their adversarial robustness. In our future work, we expect to see the successful application of this framework on more learning systems.


\balance
\bibliographystyle{IEEEtranN}	
\bibliography{IEEEabrv,RobustRS}
\end{document}